\documentclass[12pt]{article}
\usepackage{amsfonts}
\usepackage{latexsym}
\usepackage{amsmath}
\usepackage{amssymb}
\usepackage{amssymb}

\newcommand{\newsection}{    
\setcounter{equation}{0}\section}
\def\appendix#1{\addtocounter{section}{1}\setcounter{equation}{0}
\renewcommand{\thesection}{\Alph{section}}
\section*{Appendix \thesection\protect\indent \parbox[t]{11.15cm}{#1}}
\addcontentsline{toc}{section}{Appendix \thesection\ \ \ #1}}

\newcommand{\be}{\begin{eqnarray}}
\newcommand{\ee}{\end{eqnarray}}
\newcommand{\bea}{\begin{eqnarray}}
\newcommand{\eea}{\end{eqnarray}}
\newcommand{\ba}{\begin{array}}
\newcommand{\ea}{\end{array}}

\newenvironment{frcseries}{\fontfamily{frc}\selectfont}{}

\begin{document}
\begin{titlepage}
\begin{center}
\vspace{5.0cm}

\vspace{3.0cm} {\Large \bf Horizons in de-Sitter Supergravity}
\\
[.2cm]

\vspace{1.5cm}
 {\large  J. Grover$^1$ and J. Gutowski$^2$}

\vspace{0.5cm}
${}^1$ DAMTP, Centre for Mathematical Sciences\\
University of Cambridge\\
Wilberforce Road\\
Cambridge, CB3 0WA, UK\\

\vspace{0.5cm}
${}^2$ Department of Mathematics\\
King's College London\\
Strand\\
London WC2R 2LS, UK\\

\vspace{0.5cm}

\end{center}
{}
\vskip 3.0 cm
\begin{abstract}
We classify all pseudo-supersymmetric extremal near-horizon geometries in minimal five-dimensional
de-Sitter supergravity. It is shown that the only such near-horizon geometry is the near-horizon geometry of
the de-Sitter BMPV solution, and hence there are no regular extremal pseudo-supersymmetric asymptotically
de-Sitter black rings.

\end{abstract}
\end{titlepage}



\newsection{Introduction}

The existence of black ring solutions in higher dimensional supergravity theories
\cite{ring1, ring2, ring3, ring4} is interesting, as these objects violate black hole uniqueness theorems.
In particular, there are examples of supersymmetric, asymptotically flat black ring solutions with
regular horizons for which the number of free parameters in the solution exceed the number of asymptotic
conserved charges. These black ring solutions can be obtained from certain types of intersecting M2 and M5-brane
configurations in eleven dimensions. Issues of regularity of more general types of black ring solution, with arbitrary cross-sections, have been discussed in \cite{ring5}, and examples of multi-ring solutions have been found in \cite{ring6}.

The status of regular, asymptotically anti-de-Sitter black rings is considerably less clear. Attempts to construct
regular, supersymmetric, asymptotically $AdS_5$ black ring solutions, using the classification constructed in
\cite{class2} have yet to succeed, and it has been argued that such solutions cannot exist \cite{adsring1};
although examples of regular, supersymmetric, anti-de-Sitter black hole solutions do exist \cite{adsbh, pope},
see also \cite{kunduribh, herdeirobh}.
The analysis of supersymmetric black hole geometries in minimal ungauged five-dimensional supergravity
was first constructed in  \cite{reallbh}. Following this work, an analysis of supersymmetric black hole
solutions in minimal anti-de-Sitter supergravity was attempted in \cite{adsbh}. However,
in contrast with the ungauged theory, the analysis of the near-horizon geometries did not
produce a set of conditions sufficient to explicitly determine all of the possible near-horizon solutions.
Nor was it possible to adapt the global analysis of \cite{reallbh} to the case of black hole solutions of
the anti-de-Sitter theory. This is because black hole solutions of the ungauged minimal five-dimensional
supergravity can be written as fibrations over a hyper-K\"ahler base space, whereas the base space
for the anti-de-Sitter black hole solutions is only K\"ahler.
However, a detailed analysis of the possible near horizon geometries of supersymmetric anti-de-Sitter
black holes   which admit two commuting rotational isometries was constructed in \cite{kunduri1, kunduri2},
and it was shown that there are no regular black ring solutions, at least with these symmetries. So, if regular, supersymmetric, anti-de-Sitter rings exist, they will exhibit less symmetry than their asymptotically flat counterparts.

The purpose of this paper is to classify the near-horizon geometries of extremal black holes,
and investigate the status of black rings,
in five-dimensional minimal de-Sitter supergravity. 
This theory is gravity, with a positive cosmological
constant, coupled to Maxwell and Chern-Simons terms. 
By choosing appropriate coefficients in the bosonic action, one obtains a theory
for which there exists ``pseudo-supersymmetric"
solutions. Such solutions possess a (pseudo) Killing spinor, which satisfies a
Killing spinor equation. In this case, the coefficients in the
bosonic action are determined by requiring that the bosonic field equations are 
consistent with the integrability conditions of the Killing spinor equation, and with this choice 
the coefficient of the Chern-Simons term is non-zero.
In the analysis presented in this paper,
we shall consider only the conditions imposed by the Killing spinor equation; it turns out that
all of the solutions we find automatically satisfy the Einstein and gauge field equations.
We remark that although de-Sitter supergravity cannot be obtained via a truncation of type IIB supergravity, as the
corresponding anti-de-Sitter theory can be \cite{chamblin}, it can be related to
the ${\rm IIB}^{*}$ theory \cite{sabra4} which can be obtained from IIA theory via a timelike T-duality
\cite{hull}.

A classification of all pseudo-supersymmetric solutions of five-dimensional minimal
de-Sitter supergravity was constructed in
\cite{herdeiro1, herdeiro2}, and there are a number of interesting pseudo-supersymmetric black hole solutions.
The cosmological black holes found in \cite{london}, and later generalized
in \cite{cvetic2, sabra1, sabra2, sabra3} describe the generalization of the  BMPV black hole
solution \cite{bmpv} of minimal ungauged supergravity to the de-Sitter theory.
Remarkably, and in contrast to the anti-de-Sitter theory, there exist regular
multi-centre rotating de-Sitter BMPV black hole solutions.
In this sense, the de-Sitter theory has solutions which are more similar to those of
the ungauged theory than the anti-de-Sitter theory, for which there are no known regular
supersymmetric multi-centre black hole solutions. 
For non-rotating asymptotically de-Sitter or asymptotically flat  BMPV solutions
the contribution from the Chern-Simons term vanishes. However, for rotating multi-centre solutions,
the precise value of the Chern-Simons coefficient becomes important, and when the coefficient
is chosen to be consistent with supersymmetry or pseudo-supersymmetry, the multi-centre rotating solutions
in both the ungauged and de-Sitter theories can be constructed
using a superposition of harmonic functions on $\mathbb{R}^4$. 

Motivated by the black hole uniqueness theorem constructed for the ungauged theory in \cite{reallbh}, and the partial
analysis undertaken for the anti-de-Sitter theory \cite{adsbh}, we wish to determine how strongly pseudo-supersymmetry constrains the near-horizon geometries in the de-Sitter theory. 
It is by no means a priori obvious how strongly the near-horizon geometries are constrained by pseudo-supersymmetry.
This is because although the de-Sitter BMPV solutions share many properties with the asymptotically flat BMPV solution,
the ``timelike" class of pseudo-supersymmetric solutions, i.e. those for which the 1-form Killing spinor 
bilinear is timelike, and within which the de-Sitter BMPV solutions are known to lie, are written as
fibrations over a hyper-K\"ahler with torsion (HKT) 4-manifold. Hence, a naive comparison with the ungauged theory
would seem to suggest that for the de-Sitter theory, the conditions imposed by pseudo-supersymmetry might be weaker, 
as the base-space geometry is required to be hyper-K\"ahler for the ungauged theory. Further motivation for our near-horizon analysis is
provided by the recent discovery of a pseudo-supersymmetric de-Sitter black ring solution \cite{chu},
although this solution does not have a regular horizon. Hence, the question arises as to whether
there exist pseudo-supersymmetric de-Sitter black ring solutions with regular horizons.
In particular, one might expect it to be easier to find black ring solutions in the de-Sitter theory than in the anti-de-Sitter theory, as the expansion of the spacetime due to the cosmological constant might help to balance the ring against collapse due to gravitational self-attraction.

In this paper, we obtain a set of conditions on the near-horizon spacetime 
geometry and 2-form gauge field strength 
which are imposed by the Killing spinor equation.
We show that these conditions are sufficient to completely determine the metric and 
gauge field strength, and that the only such near-horizon geometry is that
of the de-Sitter BMPV solution. The plan of the paper is as follows. 
In section 2, we summarize some of the details of the de-Sitter
supergravity theory, and the construction of Gaussian null-coordinates used to describe extremal black hole
near-horizon geometries. In sections 3, 4 and 5, we analyse the constraints imposed on the
near-horizon geometries from the existence of a (pseudo) Killing spinor. In section 6, we
derive the near-horizon geometry of the de-Sitter BMPV solution, and compare it with the near-horizon geometries
obtained from the pseudo-supersymmetry analysis. In section 7, we present our conclusions. In appendices A and B, we
present the linear system of equations obtained from the Killing spinor equations, and list the components of
the spin connection of the near-horizon geometries.

\newsection{Supersymmetry and Near-Horizon Geometries}

In this section, we briefly summarize the details of minimal five-dimensional de-Sitter supergravity,
and of pseudo-supersymmetric near-horizon geometries which are solutions of this theory.
The bosonic action is
\begin{eqnarray}
 \mathcal{S}=\frac{1}{4\pi G}\int \left(\frac{1}{4}(^5R-{12 \over \ell^2})\star
1-\frac{1}{2}F\wedge \star F-\frac{2}{3\sqrt{3}}F\wedge F\wedge A
\right) \ ,
\end{eqnarray}
where we have used the same conventions as in the classification constructed in \cite{herdeiro1}, and
set $\chi = {2 \sqrt{3} \over \ell}$. $F=dA$ is a $U(1)$ field strength and $\ell$ is a real nonzero
constant, the metric signature is $(-,+,+,+,+)$. The
equations of motion are
\begin{eqnarray}
\label{einsteq}
^5R_{\alpha\beta}-2F_{\alpha\gamma}F_{\beta}^{\
  \gamma}+\frac{1}{3}g_{\alpha\beta}(F^2-{12 \over \ell^2})=0 \ ,
\end{eqnarray}
and
\begin{eqnarray}
\label{gaugeq}
 d\star
F+\frac{2}{\sqrt{3}}F\wedge F=0 \ ,
\end{eqnarray}
where $F^2\equiv F_{\alpha\beta}F^{\alpha\beta}$.

Pseudo-supersymmetric solutions of this theory admit
a Dirac spinor $\epsilon$ which satisfies the gravitino equation:

\begin{eqnarray}
\label{grav}
 \bigg[\partial_\mu +
{1\over4}\Omega_{\mu,}{}^{\nu_1 \nu_2}\Gamma_{\nu_1 \nu_2} - {i \over
4\sqrt{3}} F^{\nu_1 \nu_2} \Gamma_\mu \Gamma_{\nu_1 \nu_2} +{3i \over
2\sqrt{3}} F_{\mu}{}^{\nu} \Gamma_{\nu} +
{2 \sqrt{3} \over \ell} ({i\over4\sqrt{3}}\Gamma_{\mu} - {1\over2}A_{\mu}) \bigg]
\epsilon =0 \ ,
\nonumber \\
\end{eqnarray}
where $\Omega$ denotes the spin connection. The Dirac spinor $\epsilon$ can be written
in an appropriately chosen basis, given in \cite{halfnull}, with four complex components denoted
by $\lambda^1_+, \lambda^1_-, \lambda^{\bar{1}}_+, \lambda^{\bar{1}}_-$, and the Killing spinor equation
can be decomposed in this basis. The resulting linear system can be read off from the linear system
computed in \cite{halfnull}, and the details of this decomposition are given in Appendix A.
We work with a null basis in which
the metric is:
\begin{eqnarray}
\label{nullbasis}
ds^2 = -2 {\bf{e}}^+ {\bf{e}}^- + ({\bf{e}}^1)^2 +2 {\bf{e}}^2 {\bf{e}}^{\bar{2}} \ ,
\end{eqnarray}
where ${\bf{e}}^+, {\bf{e}}^-, {\bf{e}}^1$ are real, and ${\bf{e}}^2$, ${\bf{e}}^{\bar{2}}$ are
complex conjugate.

\subsection{Gaussian Null Co-ordinates}

In order to investigate the near-horizon geometries of extremal black holes, one
first introduces Gaussian null co-ordinates adapted to the event horizon of the
black hole. It has been shown \cite{symm3} that for extremal black hole solutions
of various higher-dimensional supergravity theories satisfying certain conditions, an
event horizon of a rotating extremal black hole must be a Killing horizon, and
the solution admits a rotational isometry. However, here we shall simply
assume that the event horizon is a Killing horizon, associated
with a timelike Killing vector ${\partial \over \partial u}$, which becomes null on the horizon.

 In the case of supersymmetric black hole solutions
in the ungauged or the anti-de-Sitter minimal five-dimensional supergravity theories,
for which one assumes that there is a Killing spinor which remains regular at the horizon,
it is straightforward to show that the Killing spinor gives rise to a Killing vector
\cite{class2, class1} as a Killing spinor bilinear. When this Killing vector is timelike, one can argue
that it also must become null on the horizon, using the reasoning  given in \cite{reallbh}.
Thus one can identify the event horizon as a Killing horizon of the Killing  vector generated by
the Killing spinor, and one can construct Gaussian null co-ordinates adapted to this Killing vector.
Furthermore, the supersymmetric black holes of these theories are extremal, and moreover, it follows
from the supersymmetry analysis that the Killing vector obtained from the Killing spinor is automatically
a symmetry of the full solution.

In contrast, for the case of pseudo-supersymmetric solutions of the de-Sitter supergravity,
it has been shown, in the classification constructed in \cite{herdeiro1}, that the Killing spinor
does not produce a Killing vector as a Killing spinor bilinear. So one cannot
identify the Killing spinor bilinear with ${\partial \over \partial u}$. Hence we shall
simply assume that ${\partial \over \partial u}$ is a symmetry of the whole solution.
Furthermore, it is known that there exist asymptotically de-Sitter pseudo-supersymmetric
black holes which are generically non-extremal \cite{london, cvetic2, sabra1, sabra2, sabra3},
but which admit an extremal limit. However, here we shall restrict our analysis to the case
of extremal black holes.

In this case, Gaussian null co-ordinates $u, r, y^M$ for $M=1,2,3$ adapted to
${\partial \over \partial u}$ can be introduced following the reasoning set out in \cite{gnull}. The metric is
 \begin{eqnarray}
 \label{gnullmet}
 ds^2 = -r^2 \Delta du^2 +2 du dr + 2r h du + ds_H^2 \ ,
 \end{eqnarray}
where the horizon is at $r=0$. $ds_H^2= \gamma_{MN} dy^M dy^N$ is
the metric on spatial cross-sections of the horizon $H$, which is analytic in
$r$ and independent of $u$, and is regular at $r=0$. $h$ is a 1-form on $H$,
and $\Delta$ is a scalar on $H$, which are again analytic in $r$, and independent of $u$.
As the black hole is extremal, one can take the near-horizon limit by setting
\begin{eqnarray}
r= \epsilon {\tilde{r}}, \qquad u = \epsilon^{-1} {\tilde{u}} \ ,
\end{eqnarray}
and taking $\epsilon \rightarrow 0$. On dropping the tilde on $r, u$, the near horizon
metric is of the same form as ({\ref{gnullmet}}), but with $h, \Delta, \gamma_{MN}$ independent
of both $r$ and $u$. We  assume that the spatial cross-section of the
horizon, $H$,  equipped with metric $ds_H^2$, is compact and simply connected.
We also assume that the resulting near-horizon geometry is pseudo-supersymmetric. In fact, in what follows,
we do not assume that the black hole bulk geometry is pseudo-supersymmetric, only that the near-horizon
geometry admits a Killing spinor.

It is convenient to use the following null basis adapted to the
Gaussian null co-ordinates:
\begin{eqnarray}
\label{gnullbasis}
{\bf{e}}^+ &=& -du
\nonumber \\ 
{\bf{e}}^- &=& dr +rh -{1 \over 2} r^2 \Delta du \ ,
 \end{eqnarray}
 and take a
$u, r$-independent basis of $H$ to be given by ${\bf{e}}^1, {\bf{e}}^2,
{\bf{e}}^{\bar{2}}$, where ${\bf{e}}^1$ is real and ${\bf{e}}^2, {\bf{e}}^{\bar{2}}$ are complex conjugate;
and the metric is given by ({\ref{nullbasis}}).
The components of the spin connection, and some other conventions associated with this basis,
are listed in Appendix B.

We further assume that the components of the gauge potential $A$ remain regular
in the near-horizon limit, and therefore take
\begin{eqnarray}
A_+ &=& -{\sqrt{3} \over 2} r \Phi
\nonumber \\
A_- &=& 0
\nonumber \\
A_m &=& B_m \ ,
\end{eqnarray}
where $m,n = 1, 2, {\bar{2}}$, and $\Phi$ and $B_m$ do not depend on $u, r$.

It follows that the components of the field strength are given by
\begin{eqnarray}
F_{+-} &=& {\sqrt{3} \over 2} \Phi
\nonumber \\
F_{-m} &=& 0
\nonumber \\
F_{+m} &=&{\sqrt{3} \over 2} r \big(\partial_m \Phi - \Phi h_m\big)
\nonumber \\
F_{mn} &=& (dB)_{mn} \ .
\end{eqnarray}

\newsection{Analysis of the Killing Spinor Equations}

To proceed, we analyse the conditions imposed on the spacetime geometry and the gauge field strength
imposed by the linear system of equations listed in Appendix A. This can be read off from the results of \cite{halfnull} which
were found using spinorial geometry techniques  originally developed to analyse
higher dimensional supergravity solutions in \cite{elevend,tend}. In this section, we integrate up the
``+" and the ``-" components of the Killing spinor equation, given by equations ({\ref{eq1}})-({\ref{eq8}),
and in the following sections, we analyse a number of integrability conditions which are necessary for 
pseudo-supersymmetry.

\subsection{Analysis of equations (A.8)-(A.11)}

The analysis of the ``-" component of the Killing spinor equations, given in equations ({\ref{eq5}})-({\ref{eq8}}) is straightforward; one finds that

\begin{eqnarray}
\label{minsol} {\lambda^1_+} &=& \mu^1_+ \nonumber \\ {\lambda^{\bar{1}}_+} &=& \mu^{\bar{1}}_+ \nonumber \\
{\lambda^1_-} &=& -r \big({i \over 2 \sqrt{2}} h_1 -{1 \over \sqrt{2}}\Phi +{1
\over \sqrt{6}} (dB)_{2 {\bar{2}}}-{1 \over \sqrt{2} \ell} \big) \mu^1_+
\nonumber \\ &-&r \big(-{i \over 2} h_2 -{1 \over \sqrt{3}} (dB)_{12} \big)
\mu^{\bar{1}}_+ + \mu^1_- \nonumber \\ {\lambda^{\bar{1}}_-} &=& -r \big(-{i \over 2} h_{\bar{2}}
+{1 \over \sqrt{3}} (dB)_{1 {\bar{2}}} \big) \mu^1_+ \nonumber \\ &-& r \big(-{i
\over 2 \sqrt{2}}h_1 -{1 \over \sqrt{2}} \Phi -{1 \over \sqrt{6}}
(dB)_{2 {\bar{2}}}-{1 \over \sqrt{2} \ell} \big) \mu^{\bar{1}}_+ +
\mu^{\bar{1}}_-  \ ,
\end{eqnarray}
where $\mu^1_{\pm}, \mu^{\bar{1}}_{\pm}$ do not
depend on $r$.

\subsection{Analysis of equations (A.4)-(A.7)}

The analysis of the ``+" component of the Killing spinor equations, given in equations ({\ref{eq1}})-({\ref{eq4}}) is also straightforward. One finds that

\begin{eqnarray}
\label{musol1}
\mu^1_- &=& \sigma^1_-
\nonumber \\
\mu^{\bar{1}}_- &=& \sigma^{\bar{1}}_-
\nonumber \\
\mu^1_+ &=& \big(-{i \over 2 \sqrt{2}} h_1 +{1 \over \sqrt{2}} \Phi +{1 \over \sqrt{6}} (dB)_{2 {\bar{2}}}
-{1 \over \sqrt{2} \ell} \big) u \sigma^1_-
\nonumber \\
&+& \big({i \over 2} h_2 -{1 \over \sqrt{3}} (dB)_{12} \big) u \sigma^{\bar{1}}_- + \sigma^1_+
\nonumber \\
\mu^{\bar{1}}_+ &=& \big( {i \over 2} h_{\bar{2}} +{1 \over \sqrt{3}} (dB)_{1 {\bar{2}}} \big) u \sigma^1_-
\nonumber \\
&+& \big({i \over 2 \sqrt{2}} h_1 +{1 \over \sqrt{2}} \Phi -{1 \over \sqrt{6}} (dB)_{2 {\bar{2}}}-{1 \over \sqrt{2} \ell} \big)
u \sigma^{\bar{1}}_- + \sigma^{\bar{1}}_+ \ ,
\end{eqnarray}
where $\sigma^1_\pm, \sigma^{\bar{1}}_\pm$ do not depend on either $r$ or $u$.
On substituting ({\ref{musol1}}) back into ({\ref{minsol}}), one obtains explicitly the
$u$, $r$ dependence of the components of the Killing spinor.

There are also a number of additional algebraic conditions. These, together with
the conditions obtained from the spatial components of the Killing spinor equation along the directions of
$H$, will be examined in the following sections.
However, before proceeding further, it is useful to note that one can locally apply a $SU(2)$
gauge transformation (not depending on $u, r$) generated by $i \Gamma_{2 {\bar{2}}}$, $\Gamma_{12}+\Gamma_{1{\bar{2}}}$ and
$i(\Gamma_{12}-\Gamma_{1 {\bar{2}}})$ to set $\sigma^{\bar{1}}_-=0$, $\sigma^1_- \in {\mathbb{R}}$.
There are then two sub-cases to consider, corresponding to $\sigma^1_- \neq 0$ and $\sigma^1_-=0$.
We analyse the two cases in the next two sections.

\newsection{Solutions with $\sigma^1_- \neq 0$}

In this section we analyse the solutions for which $\sigma^1_- \neq 0$. There are two sub-cases to consider.
In the first, we show that the spatial cross-section of the horizon is a squashed $S^3$, and 
we derive explicitly the spacetime metric and 2-form gauge field strength. We also show that 
the second sub-case admits no pseudo-supersymmetric near-horizon
geometries.

For solutions with $\sigma^1_- \neq 0$, the components of the Killing spinor are given by:

\begin{eqnarray}
 \label{kspx1} {\lambda^1_+} &=& \big(-{i \over 2 \sqrt{2}}h_1 +{1 \over
\sqrt{2}} \Phi +{1 \over \sqrt{6}}(dB)_{2 {\bar{2}}} -{1 \over \sqrt{2}
\ell} \big) u \sigma^1_- + \sigma^1_+ \nonumber \\ {\lambda^{\bar{1}}_+} &=& \big({i \over 2}
h_{\bar{2}} +{1 \over \sqrt{3}} (dB)_{1 {\bar{2}}} \big) u \sigma^1_- +
\sigma^{\bar{1}}_+ \nonumber \\ {\lambda^1_-} &=& ru \big({1 \over 4} (dh)_{2 {\bar{2}}}+{ 1
\over 2} \Delta +{3i \over 4}(\partial_1 \Phi - \Phi h_1)+{3 \over 2
\ell}\Phi \big) \sigma^1_- \nonumber \\ &-& r \big({i \over 2 \sqrt{2}} h_1
-{1 \over \sqrt{2}}\Phi +{1 \over \sqrt{6}}(dB)_{2 {\bar{2}}}-{1 \over
\sqrt{2}\ell} \big) \sigma^1_+ -r \big(-{i \over 2} h_2 -{1 \over
\sqrt{3}}(dB)_{12}\big) \sigma^{\bar{1}}_+ + \sigma^1_- \nonumber \\ {\lambda^{\bar{1}}_-} &=&
ru \big({1 \over 2 \sqrt{2}} (dh)_{1 {\bar{2}}}-{3i \over 2
\sqrt{2}}(\partial_{\bar{2}} \Phi - \Phi h_{\bar{2}}) \big) \sigma^1_- -r
\big(-{i \over 2} h_{\bar{2}}+{1 \over \sqrt{3}} (dB)_{1
{\bar{2}}}\big)\sigma^1_+ \nonumber \\ &-&r \big(-{i \over 2 \sqrt{2}} h_1 -{1
\over \sqrt{2}}\Phi -{1 \over \sqrt{6}}(dB)_{2 {\bar{2}}}-{1 \over
\sqrt{2}\ell} \big) \sigma^{\bar{1}}_+ \ .
 \end{eqnarray}

There are also a number of algebraic conditions, obtained from
equations ({\ref{eq1}})-({\ref{eq8}}), which constrain
$dB, dh, B, h, \Delta$ and $\Phi$:

\begin{eqnarray}
 \label{alg1} 
 \big({1 \over \sqrt{6}}(dB)_{2 {\bar{2}}}-{1 \over
\sqrt{2}\ell}\big)^2-\big({i \over 2 \sqrt{2}}h_1-{1 \over
\sqrt{2}}\Phi\big)^2 -\big({i \over 2}h_{\bar{2}}+{1 \over
\sqrt{3}}(dB)_{1 {\bar{2}}}\big)\big({i \over 2} h_2 +{1 \over
\sqrt{3}}(dB)_{12}\big) \nonumber \\ +{1 \over 4} (dh)_{2 {\bar{2}}}+{ 1 \over 2}
\Delta +{3i \over 4}(\partial_1 \Phi - \Phi h_1)+{3 \over 2
\ell}\Phi =0 
\nonumber \\
\end{eqnarray}
\begin{eqnarray}
\label{alg2}
-{i \over 2} h_{\bar{2}} \big(\sqrt{2}\Phi+{\sqrt{2} \over \sqrt{3}} (dB)_{2 {\bar{2}}} \big)
+{1 \over \sqrt{3}}(dB)_{1 {\bar{2}}} \big(-{i \over \sqrt{2}}h_1-{\sqrt{2} \over \ell} \big)
\nonumber \\
+{1 \over 2 \sqrt{2}} (dh)_{1 {\bar{2}}}-{3i \over 2 \sqrt{2}}(\partial_{\bar{2}} \Phi - \Phi h_{\bar{2}}) =0 \ .
\end{eqnarray}

One can also obtain additional algebraic conditions from the remaining components of
the Killing spinor equation, which can be written as:

\begin{eqnarray}
\begin{pmatrix}
\delta_1 & \delta_2 \cr
-{\bar{\delta}}_2 & {\bar{\delta}}_1
\end{pmatrix}
\begin{pmatrix}
{\lambda^1_+}
\cr
{\lambda^{\bar{1}}_+}
\end{pmatrix}
=
\begin{pmatrix}
\beta_1 & \beta_2 \cr
-{\bar{\beta}}_2 & {\bar{\beta}}_1
\end{pmatrix}
\begin{pmatrix}
{\lambda^1_+}
\cr
{\lambda^{\bar{1}}_+}
\end{pmatrix}
=0 \ ,
\end{eqnarray}
where
\begin{eqnarray} \label{scase2a} \delta_1 &=&{1 \over 4} (dh)_{2 {\bar{2}}} -{1 \over
2} \Delta -{i \over 4}(\partial_1 \Phi - \Phi h_1) +{3 \over 2
\ell}\Phi +\big(-{i \over 2 \sqrt{2}} h_1 + {1 \over
\sqrt{2}}\Phi\big)^2 \nonumber \\
 &-&\big({1 \over \sqrt{6}}(dB)_{2 {\bar{2}}} -{1 \over \sqrt{2}\ell}
\big)^2 + \big({i \over 2}h_2-{1 \over \sqrt{3}}(dB)_{12} \big)
\big({i \over 2} h_{\bar{2}} -{1 \over \sqrt{3}} (dB)_{1 {\bar{2}}} \big) \ ,
\end{eqnarray}
and
\begin{eqnarray} \label{scase2b} \delta_2 &=&-{1 \over 2 \sqrt{2}} (dh)_{12}+{i
\over 2 \sqrt{2}}(\partial_2 \Phi - \Phi h_2) +{i \over 2} h_2
\big(\sqrt{2} \Phi +{\sqrt{2} \over \sqrt{3}} (dB)_{2 {\bar{2}}} \big) \nonumber \\
&+&{1 \over \sqrt{3}} (dB)_{12} \big(-{i \over \sqrt{2}} h_1 -
{\sqrt{2} \over \ell} \big) \ ,
\end{eqnarray}
and
\begin{eqnarray} \label{scase2c} \beta_1 &=& {i \over \sqrt{2}}\big({1 \over 2}
\Delta h_1 - {1\over2}\partial_1 \Delta \big) + \big(-{ 1 \over 2
\sqrt{2}}(dh)_{12}-{3i \over 2 \sqrt{2}}(\partial_2 \Phi - \Phi h_2)
\big) \big({i \over 2} h_{\bar{2}}-{1 \over \sqrt{3}}(dB)_{1{\bar{2}}} \big) \nonumber \\
&+&\big({1 \over 4}(dh)_{2 {\bar{2}}}+{3i \over 4}(\partial_1 \Phi - \Phi
h_1)+{3 \over 2 \ell}\Phi \big) \big(-{i \over 2 \sqrt{2}} h_1 +{1
\over \sqrt{2}}\Phi -{1 \over \sqrt{6}}(dB)_{2 {\bar{2}}}+{1 \over
\sqrt{2} \ell}\big) \ , \nonumber \\ 
\end{eqnarray} 
and
\begin{eqnarray} \label{scase2d} \beta_2 &=& -i\big({1 \over 2} \Delta h_2
-{1\over2}\partial_2 \Delta\big) +\big({1 \over 4} (dh)_{2 {\bar{2}}}+{3i
\over 4}(\partial_1 \Phi-\Phi h_1 )+{3 \over 2 \ell}\Phi \big)
\big({i \over 2} h_2 +{1 \over \sqrt{3}}(dB)_{12} \big) \nonumber \\ &+&
\big(-{ 1 \over 2 \sqrt{2}}(dh)_{12}-{3i \over 2
\sqrt{2}}(\partial_2 \Phi - \Phi h_2)\big) \big({i \over 2 \sqrt{2}}
h_1 +{1 \over \sqrt{2}} \Phi +{1 \over \sqrt{6}}(dB)_{2 {\bar{2}}}+{1
\over \sqrt{2}\ell} \big)  \ . \nonumber \\
 \end{eqnarray}

So there are two sub-cases. In the first, ${\lambda^1_+}={\lambda^{\bar{1}}_+}=0$; whereas in the second
$\delta_1=\delta_2=\beta_1=\beta_2=0$.

\subsection{Solutions with ${\lambda^1_+}={\lambda^{\bar{1}}_+}=0$}

For these solutions, one has

\begin{eqnarray}
\label{scase1}
-{i \over 2 \sqrt{2}} h_1 +{1 \over \sqrt{6}} (dB)_{2 {\bar{2}}} &=& 0
\nonumber \\
\Phi-{1 \over \ell} &=& 0
\nonumber \\
{i \over 2} h_{\bar{2}} +{1 \over \sqrt{3}} (dB)_{1 {\bar{2}}} &=&0
\nonumber \\
\sigma^1_+ &=&0
\nonumber \\
\sigma^{\bar{1}}_+ &=&0 \ .
\end{eqnarray}
Furthermore, on substituting these constraints back into
({\ref{alg1}}) and ({\ref{alg2}}) one finds
\begin{eqnarray} \label{hcond0} \Delta = -{3 \over \ell^2}, \qquad  (dh)_{2 {\bar{2}}}-{3i \over
\ell} h_1 =0, \qquad (dh)_{1 2}-{3i \over \ell}h_2 =0 \ .
\end{eqnarray}
Then, from ({\ref{kspx1}}) one finds
\begin{eqnarray} {\lambda^1_+} = {\lambda^{\bar{1}}_+} = {\lambda^{\bar{1}}_-} =0, \qquad {\lambda^1_-}
= \sigma^1_- \ .
\end{eqnarray}
It follows that the vector field generated from this Killing spinor
is null.

On examining the remainder of the Killing spinor equations, one
finds the conditions
\begin{eqnarray}
 -\omega_{1,12}= \omega_{2,2 {\bar{2}}}= {2i\over\sqrt{3}}(dB)_{12},
\quad \omega_{1,2 {\bar{2}}}= {-3i\over2\ell}, \quad \omega_{2,1{\bar{2}}}=
{3i \over 2 \ell}+ {2i\over\sqrt{3}}(dB)_{2{\bar{2}}}, \quad \omega_{2,12} = 0 \ ,
\nonumber \\ 
\end{eqnarray} together with
\begin{eqnarray} 
d\log\sigma^1_- = h - {2\sqrt{3}\over\ell}B \ .
\end{eqnarray}
These are sufficient to imply
\begin{eqnarray} 
\label{hgeom} 
d{\bf{e}}^m = -{3\over\ell}\star_3 {\bf{e}}^m + h\wedge {\bf{e}}^m \ .
\end{eqnarray}

Having obtained these conditions, the reasoning used to write the near-horizon solution
explicitly in appropriately adapted co-ordinates follows very closely the analysis set out
in \cite{reallbh}. However, for convenience, we repeat this analysis here.

To proceed introduce three real co-ordinates $x^A$ ($A=1,2,3$)
such that

\begin{eqnarray}
{\bf{e}}^m = {{\bf{e}}^m}_A dx^A , \qquad h = h_{A} dx^A  \ .
\end{eqnarray}
Next note that ({\ref{hgeom}}) implies that the components of the  Ricci tensor of $H$
are given by
\begin{eqnarray} 
\label{R1} R_{AB} = (h^2 + {9\over 2 \ell^2})\gamma_{AB}  - h_A h_B - \nabla_{(A}h_{B)} \ . 
\end{eqnarray}
Here $h^2 = h_A h^A$, $\gamma_{AB}$ is the metric on $H$, and $\nabla$ is the Levi-Civita connection associated with $\gamma_{AB}$. Also note that equations ({\ref{hcond0}}) imply that
\begin{eqnarray} 
\label{hdiff1} d h = {3\over\ell} \star_3 h, \quad d \star_3 h = 0 \ , 
\end{eqnarray}
and therefore
\begin{eqnarray} 
\label{R2} R_{AB}h^B - {9\over\ell^2}h_A - \nabla^2 h_A  = 0 \ . 
\end{eqnarray}

Next consider
\begin{eqnarray} 
I = \int_H \nabla_{(A}h_{B)}\nabla^{(A}h^{B)} \ .
 \end{eqnarray}
Integrating by parts and making use of ({\ref{R2}}) we find
\begin{eqnarray} 
I = \int_H ({9\over\ell^2}h^2 - 2R_{AB}h^A h^B) \ . 
\end{eqnarray}
Substituting ({\ref{R1}}) into this expression we find, after some manipulation, that $I = 0$. Hence
\begin{eqnarray} 
\nabla_{(A}h_{B)} = 0 \ .
\end{eqnarray}

This implies that, if non-vanishing, $h$ is a Killing vector field on $H$ satisfying
\begin{eqnarray}
\nabla_A h_B = {3\over2\ell}\epsilon_{ABC}h^C \ ,
\end{eqnarray}
where $\epsilon$ is the volume form of $H$, and in particular that $h^2$ is constant on $H$. In the case that $h=0$, we find from ({\ref{R1}}) that
\begin{eqnarray} 
R_{AB} = {9\over\ell^2}\gamma_{AB} \ , 
\end{eqnarray}
and so we identify $H$ with $S^3$, taking the usual round metric. In the case that $h \neq 0 $ define
\begin{eqnarray}
 \hat{x}^m = \hat{h}^A{\bf{e}}^m_A \ ,
  \end{eqnarray}
where $\hat{h} = {h\over\sqrt{h^2}}$. These will then satisfy $\hat{x}^m\hat{x}^m = 1$, together with
\begin{eqnarray} 
\nabla_A\hat{x}^m = -h^2{\bf{e}}^m_A - {3\over\ell}\epsilon^{B}{}_{AC} h^C {\bf{e}}^m_B + h_A h^C {\bf{e}}^m_C \ . 
\end{eqnarray}

It is then straightforward to show that ${\cal{L}}_h \hat{x}^m = 0 $, and
\begin{eqnarray} 
(d\hat{x}^m d\hat{x}^m)_{AB} = ({9\over\ell^2} + h^2)(\gamma_{AB} - \hat{h}_A\hat{h}_B) \ . 
\end{eqnarray}
Since the $\hat{x}^m$ are preserved along the integral curves of $h$, 
it will be natural to use them together with a parameter along the integral curves as local coordinates on $H$. As such we can define
\begin{eqnarray} 
\hat{x}^1 = -\cos{\phi}\sin{\theta}, \qquad \hat{x}^2 = \sin{\phi}\sin{\theta}, \qquad \hat{x}^3 = \cos{\theta} \ , 
\end{eqnarray}
and introduce a parameter $\psi$ along the integral curves of $h$, normalized so that
\begin{eqnarray} 
h = -4j\mu^{-5/2}(1 - {j^2\over\mu^3})^{-{1\over2}}{\partial\over\partial\psi} \ . 
\end{eqnarray}
Here the constants $j,\mu$ are defined by
\begin{eqnarray} 
\mu = {4\over({9\over\ell^2} + h^2)}, \qquad j = \pm {8\sqrt{h^2}\over({9\over\ell^2} + h^2)^2} \ .
\end{eqnarray}
As a 1-form $h$ is then given by
\begin{eqnarray} 
h = -j\mu^{-3/2}(1 - {j^2\over\mu^3})^{{1\over2}}(d\psi + G) \ , 
\end{eqnarray}
where $G$ is a 1-form defined over $H$. We can use ({\ref{hdiff1}}) to constrain $G$, which can be chosen to take the form
\begin{eqnarray}
G = \cos{\theta}d\phi \ .
\end{eqnarray}
This fixes the volume form on $H$ to be $ \epsilon_H = {1\over8}{\ell\over|\ell|}\mu^{3/2}(1-{j^2\over\mu^3})^{1/2}\sin{\theta} d\theta \wedge d\psi \wedge d\phi$. The metric is then given by
\begin{eqnarray}
\label{finsol1}
ds^2 = {3\over\ell^2}r^2du^2 + 2du dr - 2rj\mu^{-3/2}(1 - {j^2\over\mu^3})^{{1\over2}}(d\psi + \cos{\theta}d\phi)du \nonumber \\ + {\mu\over4}[(1 - {j^2\over\mu^3})(d\psi + \cos{\theta}d\phi)^2 + d\theta^2 + \sin^2{\theta}d\phi^2] \ ,
\end{eqnarray}
where $\mu$ and $j$ satisfy
\begin{eqnarray}
{9 \over 4 \ell^2} \mu^4 - \mu^3 + j^2 =0 \ .
\end{eqnarray}
We also find that the field strength can be expressed as
\begin{eqnarray}
\label{finsol2}
F = {\sqrt{3} \over 2\ell} dr \wedge du + {\sqrt{3}\over4}{\ell\over|\ell|}{j\over\mu}\sin{\theta} d\theta \wedge d\phi \ .
\end{eqnarray}

\subsection{Solutions with $\delta_1=\delta_2=\beta_1=\beta_2=0$}

For these solutions, note that ({\ref{alg1}}), ({\ref{scase2a}})
imply that
\begin{eqnarray}
\label{screl1}
 \Phi=0 \ , \end{eqnarray}
and
\begin{eqnarray}
\label{dual1} (dB)_{2 {\bar{2}}}=0 \ .
\end{eqnarray}
On substituting these conditions into ({\ref{alg2}}) and
({\ref{scase2b}}) one then finds
\begin{eqnarray}
(dB)_{12}= (dh)_{12}= (dh)_{2{\bar{2}}} = 0 \ ,
\end{eqnarray}
as well as
\begin{eqnarray}
\label{hcond1}
d\log\Delta = h \ ,
\end{eqnarray}
from ({\ref{scase2c}}) and ({\ref{scase2d}}). These conditions set
$F = 0$. Next, substituting all these conditions back into
({\ref{alg1}}), one finds
\begin{eqnarray}
\label{hcond2}
\Delta = -{1 \over  \ell^2} - {1 \over 4}h^2  \ .
\end{eqnarray}
An analysis of the spatial components of the Killing spinor equations gives
\begin{eqnarray}
d\log\sigma^1_- &=& -{h\over4} + {\sqrt{3}B\over\ell} \ ,
\nonumber \\
d\log\sigma^1_+ &=& {h\over4} + {\sqrt{3}B\over\ell} - {i\over\ell}{\bf{e}}^1 + {\sqrt{2}i\over\ell}{\sigma^{\bar{1}}_+\over\sigma^1_+}{\bf{e}}^{{\bar{2}}} \ ,
\nonumber \\
d\log\sigma^{{\bar{1}}}_+ &=& {h\over4} + {\sqrt{3}B\over\ell} + {i\over\ell}{\bf{e}}^1 + {\sqrt{2}i\over\ell}{\sigma^1_+\over\sigma^{\bar{1}}_+}{\bf{e}}^{2} \ ,
\end{eqnarray}
together with
\begin{eqnarray}
\omega_{1,12} = \omega_{2,12} = \omega_{2,2{\bar{2}}} = 0, \quad \omega_{1,2{\bar{2}}} = -\omega_{2,1{\bar{2}}} = -{i\over\ell} \ ,
\end{eqnarray}
and
\begin{eqnarray}
\label{hcond3}\nabla_i h_j - {1\over2}h_i h_j - {2\over\ell^2}\delta_{ij} = 0 \ .
\end{eqnarray}
On taking the trace of ({\ref{hcond3}}) one obtains
\begin{eqnarray}
\nabla^i h_i = {1 \over 2} h^2 +{6 \over \ell^2} \ .
\end{eqnarray}
On integrating this expression over $H$, the integral of the LHS vanishes,
whereas the RHS is positive. Hence there can be no solutions in this class.

\newsection{Solutions with $\sigma^1_- =0$}

In this section, we analyse the solutions for which $\sigma^1_-=0$. Once more, there are several 
sub-cases to consider. In a number of these, we shall show that there can be no pseudo-supersymmetric
near-horizon geometries. In the remaining cases, we prove that the conditions imposed by
pseudo-supersymmetry are sufficient to imply that the spatial cross-section of the horizon 
is either a squashed $S^3$, or a round $S^3$,
and we obtain the spacetime metric and 2-form gauge field strength. We show that the solution 
with a squashed $S^3$ horizon cross-section is identical to that found in the previous section;
and the solution with a round $S^3$ cross-section corresponds to a special case of the same solution 
with the angular momentum set to zero.

For solutions with $\sigma^1_- =0$, one can locally apply a $SU(2)$
gauge transformation (not depending on $u, r$) generated by $i \Gamma_{2 {\bar{2}}}$, $\Gamma_{12}+\Gamma_{1{\bar{2}}}$ and
$i(\Gamma_{12}-\Gamma_{1 {\bar{2}}})$ to set $\sigma^{\bar{1}}_+=0$, $\sigma^1_+ \in {\mathbb{R}}$
(while retaining $\sigma^1_-=\sigma^{\bar{1}}_-=0$).

The components of the Killing spinor are then given by:

\begin{eqnarray}
{\lambda^1_+} &=& \sigma^1_+
\nonumber \\
{\lambda^{\bar{1}}_+} &=&0
\nonumber \\
{\lambda^1_-} &=& -r \big({i \over 2 \sqrt{2}} h_1 - {1 \over \sqrt{2}}\Phi +{1 \over \sqrt{6}}(dB)_{2 {\bar{2}}}
-{1 \over \sqrt{2} \ell} \big) \sigma^1_+
\nonumber \\
{\lambda^{\bar{1}}_-} &=& -r \big(-{i \over 2} h_{\bar{2}} +{1 \over \sqrt{3}} (dB)_{1 {\bar{2}}} \big) \sigma^1_+ \ .
\end{eqnarray}

Furthermore, by making use of a $U(1)$ gauge transformation
(independent of $u,r$), which re-scales the Killing spinor by a real ($u,r$-independent) function, one can also,
without loss of generality for the local analysis, take $\sigma^1_+=1$.
The $+$ components of the Killing spinor equation then impose the following additional algebraic constraints:

\begin{eqnarray} 
\label{algy1} 
{1 \over 4} (dh)_{2 {\bar{2}}} -{1 \over 2} \Delta -{i
\over 4}(\partial_1 \Phi - \Phi h_1) +{3 \over 2 \ell} \Phi +
\big(-{i \over 2 \sqrt{2}} h_1 +{1 \over \sqrt{2}} \Phi \big)^2 \nonumber \\
-\big({1 \over \sqrt{6}} (dB)_{2 {\bar{2}}} -{1 \over \sqrt{2} \ell}
\big)^2 +\big({i \over 2} h_2 -{1 \over \sqrt{3}} (dB)_{12} \big)
\big({i \over 2} h_{\bar{2}} -{1 \over \sqrt{3}} (dB)_{1 {\bar{2}}}\big)=0 \end{eqnarray}

\begin{eqnarray}
\label{algy2}
{1 \over 2 \sqrt{2}} (dh)_{1 {\bar{2}}}+{i \over 2 \sqrt{2}} (\partial_{\bar{2}} \Phi - \Phi h_{\bar{2}})
+{i \over 2} h_{\bar{2}} \big(\sqrt{2} \Phi -{\sqrt{2} \over \sqrt{3}}(dB)_{2 {\bar{2}}} \big)
\nonumber \\
+{1 \over  \sqrt{3}} (dB)_{1 {\bar{2}}}\big(-{i \over \sqrt{2}} h_1 +{\sqrt{2} \over \ell} \big)=0
\end{eqnarray}

\begin{eqnarray} 
\label{algy3}
 {i \over \sqrt{2}} \big({1 \over 2}\Delta h_1-
{1\over2}\partial_1 \Delta \big) +\big(-{1 \over 2
\sqrt{2}}(dh)_{12}-{3i \over 2 \sqrt{2}}(\partial_2 \Phi-\Phi h_2)
\big) \big({i \over 2} h_{\bar{2}}-{1 \over \sqrt{3}}(dB)_{1 {\bar{2}}} \big)
\nonumber \\ + \big({1 \over 4} (dh)_{2 {\bar{2}}}+{3i \over 4} (\partial_1 \Phi -
\Phi h_1)+{3 \over 2 \ell} \Phi\big) \big(-{i \over 2 \sqrt{2}}
h_1+{1 \over \sqrt{2}}\Phi -{1 \over  \sqrt{6}}(dB)_{2 {\bar{2}}}+{1 \over
\sqrt{2}} \ell \big)=0 
\nonumber \\
\end{eqnarray}

\begin{eqnarray} 
\label{algy4} 
-i \big({1 \over 2} \Delta h_{\bar{2}} -{1\over2}
\partial_{\bar{2}} \Delta\big) +\big(-{1 \over 4} (dh)_{2 {\bar{2}}}-{3i \over
4}(\partial_1 \Phi - \Phi h_1)+{3 \over 2 \ell}\Phi\big) \big({i
\over 2} h_{\bar{2}}-{1 \over \sqrt{3}}(dB)_{1 {\bar{2}}}\big) \nonumber \\ +\big({1
\over 2 \sqrt{2}}(dh)_{1 {\bar{2}}}-{3i \over 2\sqrt{2}}(\partial_{\bar{2}}
\Phi-\Phi h_{\bar{2}})\big) \big(-{i \over 2 \sqrt{2}} h_1 +{1 \over
\sqrt{2}}\Phi -{1 \over \sqrt{6}}(dB)_{2 {\bar{2}}}+{1 \over \sqrt{2}\ell}
\big) =0 \ .
\nonumber \\
\end{eqnarray}

Next consider the equations ({\ref{eq9}}), ({\ref{eq13}}), ({\ref{eq17}}).
These imply that
\begin{eqnarray}
\label{algy5}
(dB)_{2 {\bar{2}}} = -2 \sqrt{3} i \big({1 \over 4} h_1 +{\sqrt{3} \over \ell} B_1 \big) \ ,
\end{eqnarray}
and
\begin{eqnarray}
\label{algy6}
(dB)_{1 {\bar{2}}} &=& -2 \sqrt{3} i \big({1 \over 4} h_2 + {\sqrt{3} \over \ell} B_2 \big) \ .
\end{eqnarray}
Together, ({\ref{algy5}}) and ({\ref{algy6}}) are equivalent to
\begin{eqnarray}
\label{algy7}
dB = -2 \sqrt{3} \star_3 \big({1 \over 4} h +{\sqrt{3} \over \ell} B \big) \ ,
\end{eqnarray}
where $\star_3$ denotes the Hodge dual on $H$.

To proceed, substitute ({\ref{algy7}}) into the equations
({\ref{eq11}}), and the sum of ({\ref{eq15}}) with the complex conjugate of ({\ref{eq19}}).
After some manipulation, one obtains
\begin{eqnarray}
\label{algy8}
h+ 2 \sqrt{3} (\Phi+2 \ell^{-1})B + 2 \sqrt{3} \star_3 (h \wedge B)=0 \ .
\end{eqnarray}
It follows that $h$, $B$ cannot be linearly independent.

There are then a number of cases to consider, in which
$h=0, B \neq 0$; or $B=0, h \neq 0$; or $B=h=0$; or  $B \neq 0,
h \neq 0$. In the cases where either $B$ or $h$ vanish, but not
both, we find:

\begin{itemize}

\item[(i)] $h=0, B \neq 0$. Then ({\ref{algy8}}) implies that $\Phi=-2 \ell^{-1}$.
On substituting these conditions into ({\ref{algy1}}), ({\ref{algy2}}) one finds that
$dB=0$. However, ({\ref{algy7}}) then implies that $B=0$, in contradiction to the assumption that
$B \neq 0$.

\item[(ii)] $B=0, h \neq 0$. This is inconsistent with ({\ref{algy8}}).
\end{itemize}

\subsection{Solutions with $h = B = 0$}

Then ({\ref{algy1}}), ({\ref{algy2}}) imply that $\Phi$, $\Delta$
are constant, and are constrained by

\begin{eqnarray} 
\label{algaux1} \Delta = \Phi^2+{3 \over \ell} \Phi-\ell^{-2} \ .
\end{eqnarray}
Furthermore, ({\ref{algy3}}) implies that 
\begin{eqnarray}
 \Phi (\Phi+\ell^{-1})=0 \ .
\end{eqnarray}
We therefore have two subcases to consider in which either $\Phi =
0$, or $\Phi = -{1\over\ell}$.

\subsubsection{Solutions with $h = B = 0$, and $\Phi = 0$}

For this case we have $A = F = 0$, and
({\ref{algaux1}}) implies
\begin{eqnarray}
\Delta = -{1\over\ell^2} \ .
\end{eqnarray}
However, it is then straightforward to show that the remaining spatial components of the Killing
spinor equations admit no solution. There are therefore no solutions in this class.

\subsubsection{Solutions with $h = B = 0$, $ \Phi = -{1\over\ell}$}

Here we find that $A = {\sqrt{3}\over2\ell}r{\bf{e}}^+$. From
({\ref{algaux1}}) we also find that
\begin{eqnarray}
\Delta = -{3\over\ell^2} \ ,
\end{eqnarray}
and from the spatial components of the Killing spinor equation we obtain the conditions
\begin{eqnarray} 
\omega_{1,12} = \omega_{2,2{\bar{2}}} = \omega_{2,12} = 0, \quad
\omega_{1,2 {\bar{2}}} = -\omega_{2,1{\bar{2}}} = {3i\over2\ell} \ .
\end{eqnarray}

It follows that
\begin{eqnarray}
d {\bf{e}}^1 = {3i \over \ell} {\bf{e}}^2 \wedge {\bf{e}}^{\bar{2}}, \qquad d {\bf{e}}^2 = {3i \over \ell} {\bf{e}}^1 \wedge {\bf{e}}^2 \ .
\end{eqnarray}
Hence, $H$ is a 3-sphere, and one can introduce local co-ordinates $\theta, \phi, \psi$ such that
\begin{eqnarray}
\label{threesph1}
ds_H^2 ={\ell^2 \over 9} \big(d \psi^2 + d \theta^2 + d \phi^2 +2 \cos \theta d \phi d \psi \big) \ .
\end{eqnarray}
The solution is then given by
\begin{eqnarray}
ds^2 &=& {3 \over \ell^2} r^2 du^2 + 2 du dr + {\ell^2 \over 9} \big(d \psi^2 + d \theta^2 + d \phi^2 +2 \cos \theta d \phi d \psi \big) \ , 
\nonumber \\
F &=&- {\sqrt{3} \over 2 \ell} dr \wedge du \ .
\end{eqnarray}
This solution is identical to that given in ({\ref{finsol1}}), ({\ref{finsol2}}), with $j =0$, under
the replacement $\ell \rightarrow -\ell$.

\subsection{Solutions with $h \neq 0,  B \neq 0$}

In the case for which neither $B$, nor $h$ vanish, we find that
({\ref{algy8}}) implies that
\begin{eqnarray}
\label{algy9}
h=-2 \sqrt{3} (\Phi+2 \ell^{-1})B \ ,
\end{eqnarray}
and furthermore, ({\ref{algy7}}) can be rewritten as
\begin{eqnarray}
\label{algy10} dB = 3 \Phi \star_3 B \ .
\end{eqnarray}

Next, substitute ({\ref{algy9}}) and ({\ref{algy10}}) back into
({\ref{algy1}})-({\ref{algy4}}), which can then be rewritten as

\begin{eqnarray}
\label{algy11}
\Delta = {3 \over \ell}\Phi + \Phi^2 -
12\ell^{-1} (\Phi+\ell^{-1})B^2-{1 \over  \ell^2} \ ,
\end{eqnarray}
and
\begin{eqnarray}
\label{algy12}
dh = \star_3 (d \Phi-2 \sqrt{3} \Phi(\Phi+4
\ell^{-1}) B) \ ,
\end{eqnarray}
and
\begin{eqnarray}
\label{algy13}
{1 \over 2}\Delta h - {1 \over 2}d \Delta
+(\Phi+\ell^{-1})d \Phi +\sqrt{3}\Phi (\Phi^2+2 \ell^{-1}\Phi+4
\ell^{-2} )B +{2 \sqrt{3} \over \ell} \star_3(B \wedge d\Phi)=0 \ ,
\nonumber \\
\end{eqnarray}
and
\begin{eqnarray}
\label{algy14}
-2 \sqrt{6} {\cal{L}}_B \Phi -6
\sqrt{2}\Phi(\Phi+\ell^{-1})B^2+{3 \over \sqrt{2}}
\Phi(\Phi+\ell^{-1})=0 \ .
\end{eqnarray}

To proceed, take the exterior derivative of ({\ref{algy9}}) and use ({\ref{algy10}}) and ({\ref{algy12}}) to
eliminate $dh$ and $dB$. One finds
\begin{eqnarray}
d \Phi +4 \sqrt{3} \Phi(\Phi+\ell^{-1})B = 2 \sqrt{3} \star_3 (B \wedge d \Phi) \ .
\end{eqnarray}
On contracting with $B$, one finds
\begin{eqnarray}
{\cal{L}}_B \Phi + 4 \sqrt{3} \Phi(\Phi+\ell^{-1})B^2=0 \ ,
\end{eqnarray}
and on using ({\ref{algy14}}) to eliminate ${\cal{L}}_B \Phi$, one then finds
\begin{eqnarray}
\Phi(\Phi+\ell^{-1}) (18 \sqrt{2} B^2+{3 \over \sqrt{2}})=0 \ .
\end{eqnarray}

We therefore have two cases to consider, in which either
$\Phi=0$ or $\Phi=-\ell^{-1}$.

\subsubsection{Solutions with $h \neq 0, B \neq 0$, $ \Phi = 0$}

Here we find from ({\ref{algy10}})-({\ref{algy14}}) that
\begin{eqnarray}
\label{hcond5}
\Delta = -{1\over4}h^2 - {1\over\ell^2}, \quad h =
-{4\sqrt{3}\over\ell}B = d\log\Delta \ .
\end{eqnarray}
These conditions imply that $F = 0$. From the remainder of the
Killing spinor equations we find
\begin{eqnarray}
\omega_{1,12} = \omega_{2,12} = \omega_{2,2{\bar{2}}} = 0, \quad
\omega_{1,2{\bar{2}}} = -\omega_{2,1{\bar{2}}} = {i\over\ell} \ ,
 \end{eqnarray}
and
\begin{eqnarray}
\label{hcond6}
\nabla_ih_j - {1\over2}h_i h_j- {2\over\ell^2}\delta_{ij} = 0.
\end{eqnarray}

Note that this condition is identical to ({\ref{hcond3}}) in section (4.2). Hence, following the analysis detailed there, we find that there are no solutions in this case.

\subsubsection{Solutions with $h \neq 0, B \neq 0$, $ \Phi = -{1\over\ell}$}

In this case we find from ({\ref{algy10}})-({\ref{algy14}}) that
\begin{eqnarray}
\Delta = -{3\over\ell^2}, \quad h = -{2\sqrt{3}\over\ell}B,
\quad dB = -{3\over\ell}\star_3 B \ .
\end{eqnarray}
From the remainder of the Killing spinor equations we find
\begin{eqnarray} \omega_{2,12} = 0, \quad \omega_{1,12} = -\omega_{2,2{\bar{2}}} =
{2\sqrt{3}\over\ell}B_2, \quad \omega_{1,2{\bar{2}}} = {3i\over2\ell},
\quad \omega_{2,1{\bar{2}}} = -{3i\over2\ell} - {2\sqrt{3}\over\ell}B_1 \ .
\nonumber \\
\end{eqnarray}
These constraints are sufficient to imply
\begin{eqnarray} 
d{\bf{e}}^m = {3\over\ell}\star_3{\bf{e}}^m -
{2\sqrt{3}\over\ell}{\bf{e}}^m\wedge B \ . 
\end{eqnarray}

Proceeding in the same manner as in section 4.1 we find
\begin{eqnarray}
\nabla_{(i}h_{j)} = 0 \ ,
\end{eqnarray}
and so $h$ is a Killing vector field on $H$.
Following the analysis in section 4.1 directly we find that the metric takes the form
\begin{eqnarray}
\label{finsol3}
ds^2 = {3\over\ell^2}r^2du^2 + 2du dr + 2rj\mu^{-3/2}(1 - {j^2\over\mu^3})^{{1\over2}}(d\psi + \cos{\theta}d\phi)du \nonumber \\ + {\mu\over4}[(1 - {j^2\over\mu^3})(d\psi + \cos{\theta}d\phi)^2 + d\theta^2 + \sin^2{\theta}d\phi^2] \ ,
 \end{eqnarray}
 where constants $\mu$, $j$ satisfy
\begin{eqnarray}
{9 \over 4 \ell^2} \mu^4 - \mu^3 + j^2 =0 \ .
\end{eqnarray} 
We also find that the field strength can be expressed as
\begin{eqnarray}
\label{finsol4}
F = -{\sqrt{3} \over 2\ell} dr \wedge du + {\sqrt{3}\over4}{\ell\over|\ell|}{j\over\mu}\sin{\theta} d\theta \wedge d\phi \ .
\end{eqnarray}

Note in particular that this solution is identical to that given in ({\ref{finsol1}}), ({\ref{finsol2}}) (for $j \neq 0$) under
the replacements $\ell \rightarrow -\ell$, $j \rightarrow -j$.

\newsection{The de-Sitter BMPV Solution}

We have shown in the previous two sections that all pseudo-supersymmetric
near-horizon geometries have metric and gauge field strength given by ({\ref{finsol1}}), ({\ref{finsol2}});
and the spatial cross sections of the horizon are either a round or squashed $S^3$,
the round $S^3$ case corresponds to the solution with $j=0$. 
In this section, we examine the de-Sitter BMPV solution
\cite{london, sabra1, sabra2, sabra3}; we derive the Gaussian Null co-ordinates
for the special case when the solution is extremal, and obtain the corresponding
near-horizon geometry.
We shall show that the near-horizon geometry of the extremal de-Sitter BMPV solution 
corresponds to the solution given in ({\ref{finsol1}}), ({\ref{finsol2}}).

It is most straightforward to write the rotating
solution in the co-ordinates used in \cite{herdeiro1} (again making the replacement
$\chi= {2 \sqrt{3} \over \ell}$):
\begin{eqnarray}
ds^2 = - \big(1+{m \over \rho^2} -{2 \over \ell} t \big)^{-2} \big(dt + {J \over \rho^2} \sigma^3\big)
+  \big(1+{m \over \rho^2} -{2 \over \ell} t \big) \big(d \rho^2+ \rho^2 d \Omega^2 \big) \ ,
\end{eqnarray}
for constants $m, J$; and
\begin{eqnarray}
F = {\sqrt{3} \over 2} d \bigg(  \big(1+{m \over \rho^2} -{2 \over \ell} t \big)^{-1} \big(dt + {J \over \rho^2} \sigma^3\big) \bigg) \ ,
\end{eqnarray}
where $d\Omega^2 = {1 \over 4}\big( (\sigma^1)^2+(\sigma^2)^2+(\sigma^3)^2 \big)$ is the metric on $S^3$,
and $\sigma^i$ are the left-invariant 1-forms on $SU(2)$ which it will be convenient to write in terms of
Euler angles $\psi, \phi, \theta$ as
\begin{eqnarray}
\sigma^1 &=& - \sin \psi d \theta + \cos \psi \sin \theta d \phi
\nonumber \\
\sigma^2 &=& \cos \psi d \theta + \sin \psi \sin \theta d \phi
\nonumber \\
\sigma^3 &=& d \psi + \cos \theta d \phi \ .
\end{eqnarray}
To begin, set
\begin{eqnarray}
t = e^{t'}+{2 \over \ell}, \qquad \rho = e^{-{t' \over 2}} \rho'
\end{eqnarray}
and then drop the primes on $t', \rho'$. The solution then becomes:
\begin{eqnarray}
ds^2 =-X^{-2} \big(dt+{J \over \rho^2} \sigma^3\big)^2 + X \big((d \rho -{1 \over 2} \rho dt)^2+ \rho^2 d \Omega^2 \big) \ ,
\end{eqnarray}
and
\begin{eqnarray}
F = {\sqrt{3} \over 2} d \bigg( X^{-1} \big(dt+{J \over \rho^2} \sigma^3\big) \bigg) \ ,
\end{eqnarray}
where
\begin{eqnarray}
X = {m \over \rho^2} -{2 \over \ell} \ .
\end{eqnarray}
Next make the co-ordinate transformation
\begin{eqnarray}
dt &=& du + {2 \over \rho (X^3 \rho^6 -4 \rho^4 -4 J^2)} \big(X^3 \rho^6 -4 J^2 \pm 2 \rho^2
\sqrt{X^3 \rho^6 -4 J^2}  \big) d \rho
\nonumber \\
d \psi &=& d \psi' + c du +{8 J \rho \big(X^3 \rho^6 -4 J^2 \pm 2 \rho^2
\sqrt{X^3 \rho^6 -4 J^2}  \big) \over  (X^3 \rho^6 -4 \rho^4 -4 J^2)
(X^3 \rho^6-4J^2) }  d \rho \ ,
\end{eqnarray}
for a constant $c$ which will be fixed later. The metric then becomes
\begin{eqnarray}
\label{mmet1}
ds^2 &=& \big({1 \over 4}(1+c^2) \rho^2 X  -X^{-2}(1+{cJ \over \rho^2})^2 \big) du^2
\pm {2 X \rho^3 \over \sqrt{X^3 \rho^6 -4 J^2}} du d \rho
\nonumber \\
&+& \big({1 \over 2}c \rho^2 X -{2J \over \rho^2 X^2}(1+{cJ \over \rho^2}) \big) du (\sigma^3)'
+ {1 \over 4} \rho^2 X ((\sigma^1)^2+(\sigma^2)^2)
\nonumber \\
&+& \big({1 \over 4}\rho^2 X - {J^2 \over \rho^4
X^2}\big) ((\sigma^3)')^2 \ ,
\end{eqnarray}
where
\begin{eqnarray}
(\sigma^3)' = d \psi' + \cos \theta d \phi \ .
\end{eqnarray}
Finally, it will be convenient to define the function
\begin{eqnarray}
f(r) = \rho^2 X \ ,
\end{eqnarray}
so that
\begin{eqnarray}
\rho^2 = {\ell \over 2} (m-f), \qquad X ={2 \over \ell} {f \over m-f} \ .
\end{eqnarray}
The function $f$ is regular in $r$, and satisfies $f(0)>0$, and we set
\begin{eqnarray}
\label{ode1}
\mp {\ell \over 2} {f f' \over \sqrt{f^3-4J^2}} =2 \ ,
\end{eqnarray}
where $'={d \over dr}$. In these co-ordinates, the metric becomes
\begin{eqnarray}
\label{mmet2}
ds^2 &=& \big({1 \over 4}(1+c^2) f -{1 \over 4} \ell^2 f^{-2} (m +{2cJ \over \ell} -f)^2 \big) du^2
+2 du dr
\nonumber \\
&+& \big({1 \over 2} c f -J\ell f^{-2}  (m +{2cJ \over \ell} -f) \big) du  (\sigma^3)'
\nonumber \\
&+& {1 \over 4} f  ((\sigma^1)^2+(\sigma^2)^2) +  \big({1 \over 4}f - {J^2 \over f^2}\big) ((\sigma^3)')^2 \ .
\end{eqnarray}
In order for this metric to correspond to that obtained from the Gaussian null co-ordinate
system of an extremal black hole, we require that $r^0$ and $r^1$ coefficients of the $du^2$ term,
and the $r^0$ coefficient of the $du  (\sigma^3)'$ term should vanish.
This imposes the conditions
\begin{eqnarray}
c = -{4J \ell \over 3 f(0)^2}, \qquad m = {f(0)^3 + 8J^2 \over 3 f(0)^2}, \qquad
{9 \over 4  \ell^2} f(0)^4 - f(0)^3 + 4 J^2 =0 \ ,
\end{eqnarray}
and moreover, given these constraints, the sign in ({\ref{ode1}}) is fixed such that
\begin{eqnarray}
f'(0)=-{6 \over \ell^2} f(0) \ .
\end{eqnarray}
Then, on expanding out the components of the  metric ({\ref{mmet2}})
in powers of $r$, and on taking the near-horizon limit, one finds
\begin{eqnarray}
ds^2 = {3 \over \ell^2} r^2 du^2+ 2 dr du +{6J \over \ell f(0)} r du \sigma^3
+{1 \over 4} f(0) \big((\sigma^1)^2+(\sigma^2)^2 \big) +{9 f(0)^2 \over 16 \ell^2}
(\sigma^3)^2 \ ,
\nonumber \\
\end{eqnarray}
where the prime has now been dropped from $(\sigma^3)'$.
Furthermore, on applying the same transformations to the field strength $F$, and taking
the near-horizon limit, one finds that
\begin{eqnarray}
F = {\sqrt{3} \over 2 \ell} dr \wedge du +{\sqrt{3} J \over 2 f(0)} d \sigma^3 \ .
\end{eqnarray}
It is then straightforward to see that this solution is identical to that found in
({\ref{finsol1}}) and ({\ref{finsol2}}) under the identifications
\begin{eqnarray}
\mu = f(0), \qquad  j = -2 {\ell \over |\ell|} J \ .
\end{eqnarray}

\newsection{Conclusions}

In this paper we have classified all pseudo-supersymmetric regular near-horizon geometries
of extremal black holes in five-dimensional de-Sitter supergravity. We have shown that
the only such near-horizon geometries are those of the extremal de-Sitter BMPV solution.
In particular, there are no pseudo-supersymmetric extremal black ring solutions. It is remarkable
that, compared with the ungauged theory, the conditions imposed by pseudo-supersymmetry
are in fact {\it stronger} than those on the extremal solutions of the ungauged theory, for which
there exist black rings as well as black holes. This is particularly notable when one recalls that
when the 1-form Killing spinor bilinear is timelike, 
the base spaces of the de-Sitter solutions are HKT, i.e. are more weakly 
constrained than the hyper-K\"ahler base spaces of the black holes in the ungauged theory.
It would be interesting to construct an analysis of pseudo-supersymmetric non-extremal black hole
solutions, by investigating the conditions imposed by (pseudo) supersymmetry on the geometry described in Gaussian null co-ordinates. Such solutions would not admit a near-horizon limit, however one could still evaluate the conditions
on the fluxes and the metric by expanding out the relevant components in powers of the radial co-ordinate.

In addition, for the case of supersymmetric black holes in the ungauged theory \cite{reallbh}, it was possible to extend the
local analysis of the near horizon geometries into the bulk to prove a uniqueness theorem.
In particular, it was shown that the only supersymmetric black
hole whose near-horizon geometry is that of the BMPV black hole is the BMPV solution. A key step in this analysis is
the observation made in \cite{class1} that all supersymmetric solutions of the ungauged theory for which
the 1-form Killing spinor bilinear is timelike can be written as fibrations over a 4-dimensional hyper-K\"ahler manifold,
which for the case of the near-horizon BMPV solution is simply ${\mathbb{R}}^4$. Moreover, as the only complete
asymptotically flat hyper-K\"ahler manifold is ${\mathbb{R}}^4$ \cite{pope2}, it was argued that the base space for the full
black hole geometry is also ${\mathbb{R}}^4$. In contrast, for the de-Sitter theory, the timelike class of 
pseudo-supersymmetric solutions consist of 
fibrations over HKT manifolds \cite{herdeiro1}. In particular, there exist solutions which have base
spaces which are not conformal to hyper-K\"ahler manifolds, and are not Ricci flat. 
It would therefore be interesting to see if one can generalize the
global analysis given in \cite{reallbh} to the de-Sitter theory.
It may however be the case that there exist new black hole solutions, which nevertheless have
the same near-horizon geometry as the de-Sitter BMPV solution.
Finally, the classification of supersymmetric near-horizon geometries in the anti-de-Sitter theory remains to
be completed.

\vskip 0.5cm
\noindent{\bf Acknowledgements} \vskip 0.1cm
\noindent J. Grover thanks the Cambridge Commonwealth Trust and the Cambridge Philosophical
Society for support. 
J. Gutowski is supported by the EPSRC grant, EP/F069774/1. The authors would like to thank C. A. R. Herdeiro for useful discussions.
\vskip 0.5cm

 \setcounter{section}{0}

\appendix{The Linear System}

The Killing spinor equations adapted to a null basis have been
computed in Appendix B of \cite{halfnull}, using spinorial geometry techniques
originally developed to analyse eleven-dimensional supergravity solutions
\cite{elevend,tend}.
In particular, the components of the Killing spinor are denoted by
$\lambda^1_+, \lambda^1_-, \lambda^{\bar{1}}_+, \lambda^{\bar{1}}_-$,
which are complex spacetime functions.
We work with the null basis given in ({\ref{nullbasis}}).
In order to match the Killing spinor equations
computed in \cite{halfnull} with those of the minimal theory in the conventions
adopted in \cite{adsbh}, we make the following replacements:

\begin{eqnarray}
\chi &\rightarrow& {2 \sqrt{3} \over \ell}
\nonumber \\
\chi V_I X^I &\rightarrow& {1 \over \ell}
\nonumber \\
H &\rightarrow& {2 \over \sqrt{3}} F
\nonumber \\
\chi A &\rightarrow& {2 \over \sqrt{3} \ell} A
\end{eqnarray}
and we also re-label the basis indices as
\begin{eqnarray}
1 \rightarrow 2, \qquad {\bar{1}} \rightarrow {\bar{2}}, \qquad 2 \rightarrow 1
\end{eqnarray}
and make a sign change to the spin connection
\begin{eqnarray}
\omega_{\mu_1, \mu_2 \mu_3} \rightarrow - \omega_{\mu_1, \mu_2 \mu_3}
\end{eqnarray}
due to the signature difference between \cite{halfnull} and \cite{adsbh}.
Finally, in order to go from the theory with a negative cosmological constant to
a positive cosmological constant, we take $\ell \rightarrow i \ell$.
It is then straightforward to read off the Killing spinor equations:

\begin{eqnarray}
\label{eq1}
\partial_+ {\lambda^1_+} + \big({1 \over 2} \omega_{+,2 {\bar{2}}}
+{1 \over 2} \omega_{+,+-} -{i \over 2 \sqrt{3}} F_{+1}
-{\sqrt{3}  \over \ell} A_+ \big) {\lambda^1_+}
+ \big(-{1 \over \sqrt{2}} \omega_{+,12} +{i \over \sqrt{6}} F_{+2} \big) {\lambda^{\bar{1}}_+}
\nonumber \\
+ \big( -{i \over \sqrt{2}} \omega_{+,-1} +{\sqrt{2} \over \sqrt{3}} F_{+-}
+{1 \over \sqrt{6}} F_{2 {\bar{2}}} -{1 \over \sqrt{2} \ell} \big) {\lambda^1_-} + \big( i \omega_{+,-2}
-{1 \over \sqrt{3}} F_{12} \big) {\lambda^{\bar{1}}_-} =0
\nonumber \\
\end{eqnarray}


\begin{eqnarray}
\label{eq2}
\partial_+ {\lambda^{\bar{1}}_+} + \big({1 \over \sqrt{2}} \omega_{+,1 {\bar{2}}}
+{i \over \sqrt{6}} F_{+ {\bar{2}}} \big) {\lambda^1_+}
+ \big({1 \over 2} \omega_{+,+-} -{1 \over 2} \omega_{+,2 {\bar{2}}}+{i \over 2 \sqrt{3}} F_{+1}
-{\sqrt{3}  \over \ell}A_+ \big) {\lambda^{\bar{1}}_+}
\nonumber \\
+ \big(i \omega_{+,- {\bar{2}}} +{1 \over \sqrt{3}} F_{1 {\bar{2}}} \big) {\lambda^1_-} +
\big({i \over \sqrt{2}} \omega_{+,-1} + {\sqrt{2} \over \sqrt{3}} F_{+-}
-{1 \over \sqrt{6}} F_{2 {\bar{2}}} -{1 \over \sqrt{2} \ell} \big) {\lambda^{\bar{1}}_-} =0
\nonumber \\
\end{eqnarray}


\begin{eqnarray}
\label{eq3}
\partial_+ {\lambda^1_-} + \big({i \over \sqrt{2}} \omega_{+,+1} \big) {\lambda^1_+}
+ \big(-i \omega_{+,+2} \big) {\lambda^{\bar{1}}_+}
\nonumber \\
+ \big({1 \over 2} \omega_{+,2 {\bar{2}}} -{1 \over 2} \omega_{+,+-} +{\sqrt{3} i \over 2} F_{+1}
-{\sqrt{3}  \over \ell} A_+ \big) {\lambda^1_-} + \big( -{1 \over \sqrt{2}} \omega_{+,12}
-{\sqrt{3} \over \sqrt{2}} i F_{+2} \big) {\lambda^{\bar{1}}_-} =0
\nonumber \\
\end{eqnarray}


\begin{eqnarray}
\label{eq4}
\partial_+ {\lambda^{\bar{1}}_-} + \big(-i \omega_{+,+ {\bar{2}}} \big) {\lambda^1_+}
+ \big(-{i \over \sqrt{2}} \omega_{+,+1} \big) {\lambda^{\bar{1}}_+}
\nonumber \\
+ \big( {1 \over \sqrt{2}} \omega_{+,1 {\bar{2}}} -{\sqrt{3} \over \sqrt{2}} i F_{+ {\bar{2}}} \big)
{\lambda^1_-}
+ \big( -{1 \over 2} \omega_{+,+-} -{1 \over 2}  \omega_{+,2 {\bar{2}}}-{\sqrt{3} \over 2} i  F_{+1}
-{\sqrt{3}  \over \ell} A_+ \big) {\lambda^{\bar{1}}_-} =0
\nonumber \\
\end{eqnarray}


\begin{eqnarray}
\label{eq5}
\partial_- {\lambda^1_+} + \big( {1 \over 2} \omega_{-,+-} +{1 \over 2} \omega_{-,2 {\bar{2}}}
-{\sqrt{3} \over 2}i F_{-1} -{\sqrt{3} \over \ell} A_- \big) {\lambda^1_+}
\nonumber \\
+ \big(-{1 \over \sqrt{2}} \omega_{-,12} +{\sqrt{3} \over \sqrt{2}} i F_{-2} \big) {\lambda^{\bar{1}}_+}
+ \big( -{i \over \sqrt{2}} \omega_{-,-1} \big) {\lambda^1_-} + \big( i \omega_{-,-2} \big) {\lambda^{\bar{1}}_-} =0
\end{eqnarray}


\begin{eqnarray}
\label{eq6}
\partial_- {\lambda^{\bar{1}}_+} + \big({1 \over \sqrt{2}} \omega_{-,1 {\bar{2}}}+{\sqrt{3} \over \sqrt{2}} i F_{- {\bar{2}}}
\big) {\lambda^1_+}
+ \big( {1 \over 2} \omega_{-,+-} -{1 \over 2} \omega_{-,2 {\bar{2}}} +{\sqrt{3} \over 2}i F_{-1}
-{\sqrt{3}  \over \ell} A_- \big) {\lambda^{\bar{1}}_+}
\nonumber \\
+ \big(i \omega_{-,- {\bar{2}}} \big) {\lambda^1_-} + \big( {i \over \sqrt{2}} \omega_{-,-1} \big) {\lambda^{\bar{1}}_-} =0
\nonumber \\
\end{eqnarray}


\begin{eqnarray}
\label{eq7}
\partial_- {\lambda^1_-} + \big({i \over \sqrt{2}} \omega_{-,+1} -{\sqrt{2} \over \sqrt{3}} F_{+-}
+{1 \over \sqrt{6}} F_{2 {\bar{2}}} -{1 \over \sqrt{2} \ell} \big) {\lambda^1_+}
+ \big( -i \omega_{-,+2} -{1 \over \sqrt{3}} F_{12} \big) {\lambda^{\bar{1}}_+}
\nonumber \\
+ \big(-{1 \over 2} \omega_{-,+-} +{1 \over 2} \omega_{-,2 {\bar{2}}}+{i \over 2  \sqrt{3}} F_{-1}
-{\sqrt{3}  \over \ell} A_- \big) {\lambda^1_-}
+ \big(-{1 \over \sqrt{2}} \omega_{-,12} -{i \over \sqrt{6}} F_{-2} \big) {\lambda^{\bar{1}}_-} =0
\nonumber \\
\end{eqnarray}


\begin{eqnarray}
\label{eq8}
\partial_- {\lambda^{\bar{1}}_-}
+ \big(-i \omega_{-,+ {\bar{2}}} +{1 \over \sqrt{3}} F_{1 {\bar{2}}} \big) {\lambda^1_+}
+ \big(-{i \over \sqrt{2}} \omega_{-,+1} - {\sqrt{2} \over \sqrt{3}} F_{+-}
-{1 \over \sqrt{6}} F_{2 {\bar{2}}} -{1 \over \sqrt{2} \ell} \big) {\lambda^{\bar{1}}_+}
\nonumber \\
+ \big({1 \over \sqrt{2}} \omega_{-,1 {\bar{2}}} - {i \over \sqrt{6}} F_{- {\bar{2}}} \big) {\lambda^1_-}
+ \big(-{1 \over 2} \omega_{-,+-}-{1 \over 2} \omega_{-,2 {\bar{2}}} -{i \over 2 \sqrt{3}} F_{-1}
-{\sqrt{3}  \over \ell} A_- \big) {\lambda^{\bar{1}}_-} =0
\nonumber \\
\end{eqnarray}


\begin{eqnarray}
\label{eq9}
\partial_1 {\lambda^1_+} + \big({1 \over 2} \omega_{1,+-}
+{1 \over 2} \omega_{1,2 {\bar{2}}} +{i \over 2 \sqrt{3}} F_{+-}
+{i \over 2 \sqrt{3}} F_{2 {\bar{2}}} -{\sqrt{3} \over \ell} A_1 -{i \over 2 \ell} \big) {\lambda^1_+}
\nonumber \\
+ \big(-{1 \over \sqrt{2}} \omega_{1,12} +{\sqrt{2} \over \sqrt{3}} i F_{12} \big) {\lambda^{\bar{1}}_+}
+ \big(-{i \over \sqrt{2}} \omega_{1,-1}-{\sqrt{2} \over \sqrt{3}} F_{-1} \big) {\lambda^1_-}
\nonumber \\
+ \big(i \omega_{1,-2} -{1 \over \sqrt{3}} F_{-2} \big) {\lambda^{\bar{1}}_-} =0
\nonumber \\
\end{eqnarray}


\begin{eqnarray}
\label{eq10}
\partial_1 {\lambda^{\bar{1}}_+} + \big({1 \over \sqrt{2}} \omega_{1,1 {\bar{2}}}
+{\sqrt{2} \over \sqrt{3}} i F_{1 {\bar{2}}} \big) {\lambda^1_+}
\nonumber \\
+ \big({1 \over 2} \omega_{1,+-} -{1 \over 2} \omega_{1,2 {\bar{2}}}
-{i \over 2 \sqrt{3}} F_{+-} +{i \over 2 \sqrt{3}} F_{2 {\bar{2}}} -{\sqrt{3}  \over \ell} A_1
+{i \over 2 \ell} \big) {\lambda^{\bar{1}}_+}
\nonumber \\
+ \big(i \omega_{1,-{\bar{2}}} +{1 \over \sqrt{3}} F_{- {\bar{2}}} \big) {\lambda^1_-}
+ \big({i \over \sqrt{2}} \omega_{1,-1} -{\sqrt{2} \over \sqrt{3}} F_{-1} \big) {\lambda^{\bar{1}}_-} =0
\end{eqnarray}


\begin{eqnarray}
\label{eq11}
\partial_1 {\lambda^1_-} + \big({i \over \sqrt{2}} \omega_{1,+1} -{\sqrt{2} \over \sqrt{3}} F_{+1} \big) {\lambda^1_+}
+ \big(-i \omega_{1,+2} -{1 \over \sqrt{3}} F_{+2} \big) {\lambda^{\bar{1}}_+}
\nonumber \\
+ \big(-{1 \over 2}\omega_{1,+-} +{1 \over 2} \omega_{1,2 {\bar{2}}}
+{i \over 2 \sqrt{3}} F_{+-} -{i \over 2 \sqrt{3}} F_{2 {\bar{2}}}
-{\sqrt{3} \over \ell} A_1 +{i \over 2 \ell} \big) {\lambda^1_-}
\nonumber \\
+ \big(-{1 \over \sqrt{2}} \omega_{1,12} -{\sqrt{2} \over \sqrt{3}} i F_{12} \big) {\lambda^{\bar{1}}_-} =0
\end{eqnarray}


\begin{eqnarray}
\label{eq12}
\partial_1 {\lambda^{\bar{1}}_-} + \big(-i \omega_{1,+ {\bar{2}}} +{1 \over \sqrt{3}} F_{+ {\bar{2}}} \big) {\lambda^1_+}
+ \big(-{i \over \sqrt{2}} \omega_{1,+1} -{\sqrt{2} \over \sqrt{3}} F_{+1} \big) {\lambda^{\bar{1}}_+}
\nonumber \\
+ \big({1 \over \sqrt{2}} \omega_{1,1 {\bar{2}}} -{\sqrt{2} \over \sqrt{3}} i F_{1 {\bar{2}}} \big) {\lambda^1_-}
\nonumber \\
+ \big(-{1 \over 2} \omega_{1,+-} -{1 \over 2} \omega_{1,2 {\bar{2}}} -{i \over 2 \sqrt{3}} F_{+-}
-{i \over 2 \sqrt{3}} F_{2 {\bar{2}}} -{\sqrt{3}  \over \ell} A_1 -{i \over 2 \ell} \big) {\lambda^{\bar{1}}_-} =0
\end{eqnarray}


\begin{eqnarray}
\label{eq13}
\partial_2 {\lambda^1_+} + \big({1 \over 2} \omega_{2,+-} +{1 \over 2} \omega_{2, 2 {\bar{2}}}
+{\sqrt{3} \over 2} i F_{12} -{\sqrt{3}  \over \ell} A_2 \big) {\lambda^1_+}
\nonumber \\
+ \big( -{1 \over \sqrt{2}} \omega_{2,12} \big) {\lambda^{\bar{1}}_+} + \big(-{i \over \sqrt{2}} \omega_{2,-1}
-{\sqrt{3} \over \sqrt{2}} F_{-2} \big) {\lambda^1_-} + \big(i \omega_{2,-2} \big) {\lambda^{\bar{1}}_-} =0
\end{eqnarray}


\begin{eqnarray}
\label{eq14}
\partial_2 {\lambda^{\bar{1}}_+}
+ \big({1 \over \sqrt{2}} \omega_{2,1 {\bar{2}}} -{i \over \sqrt{6}} F_{+-} +{\sqrt{2} \over \sqrt{3}} i F_{2 {\bar{2}}}
+{i \over \sqrt{2} \ell} \big) {\lambda^1_+}
\nonumber \\
+ \big({1 \over 2} \omega_{2,+-} -{1 \over 2} \omega_{2, 2 {\bar{2}}}-{i \over 2 \sqrt{3}} F_{12}
-{\sqrt{3}  \over \ell} A_2 \big) {\lambda^{\bar{1}}_+}
\nonumber \\
+ \big(i \omega_{2,- {\bar{2}}} - {1 \over \sqrt{3}} F_{-1} \big) {\lambda^1_-}
+ \big({i \over \sqrt{2}} \omega_{2,-1} - {1 \over \sqrt{6}} F_{-2} \big) {\lambda^{\bar{1}}_-} =0
\nonumber \\
\end{eqnarray}


\begin{eqnarray}
\label{eq15}
\partial_2 {\lambda^1_-} + \big({i \over \sqrt{2}} \omega_{2,+1} - {\sqrt{3} \over \sqrt{2}} F_{+2} \big) {\lambda^1_+}
+ \big( -i \omega_{2,+2}  \big) {\lambda^{\bar{1}}_+}
\nonumber \\
+ \big(-{1 \over 2} \omega_{2,+-} +{1 \over 2} \omega_{2,2 {\bar{2}}} -{\sqrt{3} \over 2} i F_{12}
-{\sqrt{3}  \over \ell} A_2 \big) {\lambda^1_-} + \big( -{1 \over \sqrt{2}} \omega_{2,12} \big) {\lambda^{\bar{1}}_-} =0
\end{eqnarray}


\begin{eqnarray}
\label{eq16}
\partial_2 {\lambda^{\bar{1}}_-}
+ \big(-i \omega_{2,+ {\bar{2}}} - {1 \over \sqrt{3}} F_{+1} \big) {\lambda^1_+}
+ \big( -{i \over \sqrt{2}} \omega_{2,+1} - {1 \over \sqrt{6}} F_{+2} \big) {\lambda^{\bar{1}}_+}
\nonumber \\
+ \big({1 \over \sqrt{2}} \omega_{2,1 {\bar{2}}} -{i \over \sqrt{6}} F_{+-}
-{\sqrt{2} \over \sqrt{3}} i F_{2 {\bar{2}}}-{i \over \sqrt{2} \ell} \big) {\lambda^1_-}
\nonumber \\
+ \big(-{1 \over 2} \omega_{2,+-} -{1 \over 2} \omega_{2,2 {\bar{2}}} +{i \over 2 \sqrt{3}} F_{12}
-{\sqrt{3}  \over \ell} A_2 \big) {\lambda^{\bar{1}}_-}=0
\end{eqnarray}


\begin{eqnarray}
\label{eq17}
\partial_{\bar{2}} {\lambda^1_+} + \big( {1 \over 2} \omega_{{\bar{2}}, + -}+{1 \over 2} \omega_{{\bar{2}},2 {\bar{2}}}
+{i \over 2 \sqrt{3}} F_{1 {\bar{2}}} -{\sqrt{3} \over \ell} A_{\bar{2}} \big) {\lambda^1_+}
\nonumber \\
+ \big(-{1 \over \sqrt{2}} \omega_{{\bar{2}}, 1 2}-{i \over \sqrt{6}} F_{+-} -{\sqrt{2} \over \sqrt{3}}i F_{2 {\bar{2}}}
+{i \over \sqrt{2} \ell} \big) {\lambda^{\bar{1}}_+}
\nonumber \\
+ \big(-{i \over \sqrt{2}} \omega_{{\bar{2}}, -1} - {1 \over \sqrt{6}}F_{- {\bar{2}}} \big) {\lambda^1_-}
+ \big( i \omega_{{\bar{2}}, -2} +{1 \over \sqrt{3}} F_{-1} \big) {\lambda^{\bar{1}}_-} =0
\end{eqnarray}


\begin{eqnarray}
\label{eq18}
\partial_{\bar{2}} {\lambda^{\bar{1}}_+} + \big( {1 \over \sqrt{2}} \omega_{{\bar{2}},  1 {\bar{2}}} \big) {\lambda^1_+}
+ \big({1 \over 2} \omega_{{\bar{2}}, +-}-{1 \over 2} \omega_{{\bar{2}}, 2 {\bar{2}}}
-{\sqrt{3} \over 2} i F_{1 {\bar{2}}} -{\sqrt{3}  \over \ell} A_{\bar{2}} \big) {\lambda^{\bar{1}}_+}
\nonumber \\
+ \big(i \omega_{{\bar{2}}, - {\bar{2}}} \big) {\lambda^1_-} + \big({i \over \sqrt{2}} \omega_{{\bar{2}}, -1}
-{\sqrt{3} \over \sqrt{2}} F_{- {\bar{2}}} \big) {\lambda^{\bar{1}}_-} =0
\end{eqnarray}


\begin{eqnarray}
\label{eq19}
\partial_{\bar{2}} {\lambda^1_-} + \big({i \over \sqrt{2}} \omega_{{\bar{2}}, +1} -{1 \over \sqrt{6}} F_{+ {\bar{2}}} \big) {\lambda^1_+}
+ \big(-i \omega_{{\bar{2}}, + 2} +{1 \over \sqrt{3}} F_{+1} \big) {\lambda^{\bar{1}}_+}
\nonumber \\
+ \big(-{1 \over 2} \omega_{{\bar{2}} , + -} + {1 \over 2} \omega_{{\bar{2}}, 2 {\bar{2}}}-{i \over 2 \sqrt{3}} F_{1 {\bar{2}}}
-{\sqrt{3}  \over \ell} A_{\bar{2}} \big) {\lambda^1_-}
\nonumber \\
+ \big( -{1 \over \sqrt{2}} \omega_{{\bar{2}},1 2}-{i \over \sqrt{6}} F_{+-} +{\sqrt{2} \over \sqrt{3}}i F_{2 {\bar{2}}}
-{i \over \sqrt{2} \ell} \big) {\lambda^{\bar{1}}_-} =0
\end{eqnarray}


\begin{eqnarray}
\label{eq20}
\partial_{\bar{2}} {\lambda^{\bar{1}}_-} + \big(-i \omega_{{\bar{2}}, + {\bar{2}}} \big) {\lambda^1_+} + \big(-{i \over \sqrt{2}} \omega_{{\bar{2}}, +1}
-{\sqrt{3} \over \sqrt{2}} F_{+ {\bar{2}}} \big) {\lambda^{\bar{1}}_+}
\nonumber \\
+ \big({1 \over \sqrt{2}} \omega_{{\bar{2}}, 1 {\bar{2}}} \big) {\lambda^1_-} + \big(-{1 \over 2} \omega_{{\bar{2}}, +-}
-{1 \over 2} \omega_{{\bar{2}}, 2 {\bar{2}}}+{\sqrt{3} \over 2} i F_{1 {\bar{2}}} -{\sqrt{3}  \over \ell} A_{\bar{2}} \big) {\lambda^{\bar{1}}_-}
=0 \ .
\nonumber \\
\end{eqnarray}

\appendix{The Spin Connection}

In this appendix we list the components of the spin connection associated with the null
basis given in ({\ref{nullbasis}}) and ({\ref{gnullbasis}}).

Note that
\begin{eqnarray}
d {\bf{e}}^+ &=& 0 \nonumber \\ d {\bf{e}}^- &=& -r \Delta
{\bf{e}}^+ \wedge {\bf{e}}^- + {\bf{e}}^+ \wedge \big({1 \over 2}
r^2 \Delta h - r^2 {1\over2}d \Delta \big) + {\bf{e}}^- \wedge h + r
dh \ .
\end{eqnarray}

Also, if $g$ is any function, then the relationship between frame and co-ordinate indices is:
\begin{eqnarray}
\partial_+ g &=& -\partial_u g -{1 \over 2} r^2 \Delta \partial_r g
\nonumber \\
\partial_- g &=& \partial_r g
\nonumber \\
\partial_i g &=& {\tilde{\partial}}_i g -r \partial_r g h_i \ ,
\end{eqnarray}
where $i=1,2, {\bar{2}}$, where ${\tilde{\partial}}_i g$ denotes the $i$-th component of
${\tilde{d}} g$ taken with respect to the basis ${\bf{e}}^i$ on $H$, and ${\tilde{d}} g$
is the exterior derivative of $g$ with $u, r$ held constant.

The components of the spin connection are then given by

\begin{eqnarray}
\omega_{+,+-} &=& -r \Delta \nonumber \\ \omega_{+,+m} &=& r^2 ({1
\over 2} \Delta h_m - {1\over2}\partial_m \Delta) \nonumber \\ \omega_{+,-m}
&=& {1 \over 2} h_m \nonumber \\ \omega_{+,mn} &=& {1 \over 2} r (dh)_{mn}
\nonumber \\ \omega_{-,+-} &=& 0 \nonumber \\ \omega_{-,+m} &=& {1 \over 2} h_m \nonumber \\
\omega_{-,-m} &=&0 \nonumber \\ \omega_{-,mn} &=&0 \nonumber \\ \omega_{m,+-} &=& -{1
\over 2} h_m \nonumber \\ \omega_{m,+n} &=& {1 \over 2} r (dh)_{mn} \nonumber \\
\omega_{m,-n} &=&0  \ ,
\end{eqnarray}
and $\omega_{m,pq}$ are the components of the spin connection of $H$, equipped
with basis ${\bf{e}}^i$.



\begin{thebibliography}{99}

\bibitem{ring1}
H. Elvang, R. Emparan, D. Mateos and H. S. Reall,
\textit{A Supersymmetric black ring}, Phys. Rev. Lett. {\bf{93}} (2004) 211302;  hep-th/0407065.

\bibitem{ring2}
R. Emparan and H. S. Reall,
\textit{A Rotating black ring solution in five-dimensions},
Phys. Rev. Lett. {\bf{88}} (2002) 101101;  hep-th/0110260.

\bibitem{ring3}
I. Bena and N. P. Warner, \textit{One ring to rule them all ... and in the darkness bind them?}
Adv. Theor. Math. Phys. {\bf{9}} (2005) 667; hep-th/0408106.

\bibitem{ring4}
H. Elvang, R. Emparan, D. Mateos and H. S. Reall,
\textit{Supersymmetric black rings and three-charge supertubes,}
Phys. Rev. {\bf{D71}} (2005) 024033.

\bibitem{ring5}
G. T. Horowitz and H. S. Reall,
\textit{How hairy can a black ring be?}
Class. Quant. Grav. {\bf{22}} (2005) 1289; hep-th/0411268.

\bibitem{ring6}
J. P. Gauntlett and J. B. Gutowski,
\textit{Concentric black rings,}
Phys.Rev. {\bf{D71}} (2005) 025013;  hep-th/0408010. \\
J. P. Gauntlett and J. B. Gutowski,
\textit{General concentric black rings,}
Phys. Rev. {\bf{D71}} (2005) 045002;  hep-th/0408122.

\bibitem{class2}
  J. P. Gauntlett and J. B. Gutowski,
  \textit{All supersymmetric solutions of minimal gauged supergravity in five
  dimensions,}
  Phys. Rev.  D {\bf 68} (2003) 105009
  [Erratum-ibid.\  D {\bf 70} (2004) 089901]; hep-th/0304064.

\bibitem{adsring1}
M. M. Caldarelli, R. Emparan and M. J. Rodriguez,
\textit{Black Rings in (Anti)-deSitter space,}
JHEP 0811 (2008) 011; arXiv:0806.1954 [hep-th]

\bibitem{adsbh} J.~B.~Gutowski and H.~S.~Reall, \textit{Supersymmetric $AdS_{5}$
Black Holes,} JHEP 02 (2004) 006; [hep-th/0401042].

\bibitem{pope}
Z. W. Chong, M. Cvetic, H. Lu and C. N. Pope,
\textit{Five-dimensional gauged supergravity black holes with independent rotation parameters,}
Phys. Rev. {\bf{D72}} (2005) 041901;  hep-th/0505112.

\bibitem{kunduribh}
H. K. Kunduri, J. Lucietti and H. S. Reall,
\textit{Supersymmetric multi-charge AdS(5) black holes,}
JHEP 0604 (2006) 036; hep-th/0601156.

\bibitem{herdeirobh}
P. Figueras, C. A. R. Herdeiro and F. P. Correia,
\textit{On a class of 4D Kahler bases and AdS(5) supersymmetric Black Holes,}
JHEP 0611 (2006) 036; hep-th/0608201.

\bibitem{reallbh}
H. S. Reall,
\textit{Higher dimensional black holes and supersymmetry},
Phys. Rev. {\bf{D68}} (2003) 024024; hep-th/0211290.

\bibitem{kunduri1}
H. K. Kunduri, J. Lucietti and H. S. Reall,
\textit{Do supersymmetric anti-de Sitter black rings exist?}
JHEP 0702 (2007) 026; hep-th/0611351.

\bibitem{kunduri2}
H. K. Kunduri and J. Lucietti,
\textit{Near-horizon geometries of supersymmetric AdS(5) black holes,}
JHEP 0712 (2007) 015; arXiv:0708.3695 [hep-th].










\bibitem{chamblin}
A. Chamblin, R. Emparan, C. V. Johnson and R. C. Myers, \textit{Charged AdS black holes
and catastrophic holography,} Phys. Rev. {\bf{D60}} (1999) 064018; hep-th/9902170.

\bibitem{sabra4}
J. T. Liu, W. A. Sabra and W. Y. Wen, \textit{Consistent reductions of
IIB*/M* theory and de Sitter supergravity,} JHEP 0401 (2004) 007; hep-th/0304253.

\bibitem{hull}
C. M. Hull, \textit{Timelike T-duality, de Sitter space, large N gauge theories and
topological field theory,} JHEP 9807 (1998) 021; hep-th/9806146.

\bibitem{herdeiro1}
J. Grover, J. B. Gutowski, C. A. R. Herdeiro and W. Sabra,
\textit{HKT Geometry and de Sitter Supergravity,}
Nucl. Phys. {\bf{B809}} (2009) 406; arXiv:0806.2626 [hep-th]

\bibitem{herdeiro2}
J. Grover, J. B. Gutowski, C. A. R. Herdeiro,
P. Meessen, A. Palomo-Lozano and W. A. Sabra,
\textit{Gauduchon-Tod structures, Sim holonomy and De Sitter supergravity,}
JHEP 0907 (2009) 069;  arXiv:0905.3047 [hep-th].


\bibitem{london}
L. A. J. London,
\textit{Arbitrary dimensional cosmological multi - black holes,}
Nucl. Phys. {\bf{B434}} (1995) 709.

\bibitem{cvetic2}
K. Behrndt and M. Cvetic, \textit{Time-dependent backgrounds from supergravity
with gauged non-compact R-symmetry,} 
Class. Quant. Grav. {\bf{20}} (2003) 4177; hep-th/0303266.









\bibitem{sabra1}
D. Klemm and W. Sabra,
\textit{General (anti-)de Sitter black holes in five-dimensions,}
JHEP 0102 (2001) 031;  hep-th/0011016

\bibitem{sabra2}
D. Klemm and W. Sabra,
\textit{Charged rotating black holes in 5-D Einstein-Maxwell (A)dS gravity,}
Phys. Lett. {\bf{B503}} (2001) 147; hep-th/0010200.

\bibitem{sabra3}
J. T. Liu and W. Sabra,
\textit{Multicentered black holes in gauged D = 5 supergravity,}
Phys. Lett. {\bf{B498}} (2001) 123; hep-th/0010025.

\bibitem{bmpv}
J. C. Breckenridge, R. C. Myers, A. W. Peet and C. Vafa,
\textit{D-branes and spinning black holes},
Phys. Lett. {\bf{B391}} (1997) 93; hep-th/9602065.

\bibitem{chu}
C. Chu and S. Dai,
\textit{Black Ring with a Positive Cosmological Constant,}
Phys. Rev. {\bf{D75}} (2007) 064016;  hep-th/0611325.

\bibitem{halfnull}
J. Grover, Jan B. Gutowski and Wafic Sabra,
\textit{Null Half-Supersymmetric Solutions in Five-Dimensional Supergravity};
[arXiv:0802.0231 (hep-th)].

\bibitem{symm3}
S. Hollands and S. Yazadjiev,
\textit{On the `Stationary Implies Axisymmetric' Theorem for Extremal Black Holes in Higher Dimensions},
Commun. Math. Phys. {\bf{291}} (2009) 403;  arXiv:0809.2659 [gr-qc].

\bibitem{class1}
J. P. Gauntlett, J. B. Gutowski, C. M. Hull, S. Pakis and H. S. Reall,
\textit{All supersymmetric solutions of minimal supergravity in five- dimensions},
Class. Quant. Grav. {\bf{20}} (2003) 4587; hep-th/0209114.

\bibitem{gnull}
H. Friedrich, I. Racz and R. M. Wald,
\textit{On the rigidity theorem for space-times with a stationary event horizon or a compact Cauchy horizon,}
Commun. Math. Phys. {\bf{204}} (1999) 691; gr-qc/9811021.

\bibitem{elevend}
J. Gillard, U. Gran and G. Papadopoulos,
\textit{The Spinorial geometry of supersymmetric backgrounds,}
Class. Quant. Grav. {\bf{22}} (2005) 1033; hep-th/0410155.

\bibitem{tend}
U. Gran, J. Gutowski and G. Papadopoulos,
\textit{The Spinorial geometry of supersymmetric IIb backgrounds},
Class. Quant. Grav. {\bf{22}} (2005) 2453; hep-th/0501177.

\bibitem{pope2}
G. W. Gibbons and C. N. Pope,
\textit{The Positive Action Conjecture and Asymptotically Euclidean Metrics in Quantum Gravity,}
Commun. Math. Phys. {\bf{66}} (1979) 267.

































\end{thebibliography}
\end{document}